\documentclass[3p,twocolumn]{elsarticle}
\usepackage[hyperindex,breaklinks]{hyperref}
\pdfoutput=1
\biboptions{sort&compress}
\usepackage{dblfloatfix}
\usepackage{amssymb}
\usepackage{amsmath}
\usepackage{preprintcover}
\usepackage{slashed}
\usepackage{graphicx}
\usepackage{multirow}
\usepackage{atlasphysics} 

\PreprintCoverPaperTitle{Measurement of the jet radius and transverse
  momentum dependence of inclusive jet suppression in lead-lead
  collisions at $\sqrt{s_{\mathrm{NN}}}=2.76$~\TeV\ with the ATLAS detector}

\PreprintIdNumber{CERN-PH-EP-2012-134}

\PreprintCoverAbstract{
Measurements of inclusive jet suppression in heavy ion collisions at the 
LHC provide direct sensitivity to the physics of jet quenching. 
In a sample of lead-lead collisions at $\sqrts = 2.76$~\TeV\ corresponding to an 
integrated luminosity of approximately $7~\mu \mathrm{b}^{-1}$, ATLAS
has measured jets with a calorimeter system over the pseudorapidity interval
$|\eta| < 2.1$ and over the 
transverse momentum range $38 < \pT < 210$~\GeV. Jets were reconstructed using the
anti-\kt\ algorithm with values for the distance parameter that determines the 
nominal jet radius of $R = 0.2, 0.3, 0.4$ and 0.5. The centrality
dependence of the jet yield is characterized by the jet
``central-to-peripheral ratio,'' \Rcp. Jet production is found to be
suppressed by approximately a factor of two in the 10\% most central 
collisions relative to peripheral collisions.  \Rcp\ varies smoothly
with centrality as characterized by the number of participating
nucleons. The observed suppression is only weakly dependent on transverse
momentum. A significant dependence of the \Rcp\ on the jet radius is
observed for jets with  $\pt < 100$~\GeV. These results provide the first direct
measurement of inclusive jet suppression in heavy ion collisions and
complement previous measurements of dijet transverse energy imbalance
at the LHC. 
}

\PreprintJournalName{PLB}

\newcommand{\SixtyToEighty}{60\mbox{--}80}
\newcommand{\ZeroToTen}{0\mbox{--}10}

\newcommand{\pp}{\mbox{$pp$}}
\newcommand{\AuAu}{\mbox{Au+Au}}
\newcommand{\PbPb}{\mbox{Pb+Pb}}
\newcommand{\Ncoll}{\mbox{$N_{\mathrm{coll}}$}}

\newcommand{\Rcollcent}{\mbox{$R_{\mathrm{coll}}$}}
\newcommand{\Npart}{\mbox{$N_{\mathrm{part}}$}}

\newcommand{\Rcp}{\mbox{$R_{\rm CP}$}}
\newcommand{\delRcp}{\mbox{$\delta R_{\rm CP}^{\mathrm{sys}}$}}
\newcommand{\RTwo}{\mbox{$R= 0.2$}}

\newcommand{\RFour}{\mbox{$R = 0.4$}}
\newcommand{\RFive}{\mbox{$R = 0.5$}}
\newcommand{\kt}{\mbox{$k_{t}$}}
\newcommand{\antikt}{\mbox{anti-\kt}}

\newcommand{\sqrts}{\mbox{$\sqrt{s_{\mathrm{NN}}}$}}
\newcommand{\vtwo}{\mbox{$v_2$}}
\newcommand{\ETfcal}{\mbox{$\Sigma E_{\mathrm{T}}^{\mathrm{FCal}}$}}
\newcommand{\ETtrue}{\mbox{$E_{\mathrm{T}}^{\mathrm{truth}}$}}
\newcommand{\pTtrue}{\mbox{$p_{\mathrm{T}}^{\mathrm{truth}}$}}
\newcommand{\pTrec}{\mbox{$p_{\mathrm{T}}^{\mathrm{rec}}$}}
\newcommand{\ETrec}{\mbox{$E_{\mathrm{T}}^{\mathrm{rec}}$}}
\newcommand{\ETtbyf}{\mbox{$E_{\mathrm{T}}^{3\times 4}$}}
\newcommand{\ETsbys}{\mbox{$E_{\mathrm{T}}^{7\times 7}$}}
\newcommand{\DEt}{\mbox{$\Delta \ET$}}
\newcommand{\Dpt}{\mbox{$\Delta \pT$}}
\newcommand{\phat}{\mbox{$\hat{p}_{\mathrm{T}}$}}

\newcommand{\pthat}{\mbox{$\hat{p}_{\mathrm{T}}$}}
\newcommand{\pttrkjet}{\mbox{$p_{{\mathrm{T}}}^{\mathrm{trkjet}}$}}

\newcommand{\rcpcorr}{\mbox{$R_{\mathrm{CP}}$}}
\newcommand{\rcpraw}{\mbox{$R_{\mathrm{CP}}^{\mathrm{meas}}$}}
\newcommand{\centup}{^{\mathrm{cent}}}

\newcommand{\xini}{\mbox{$x_{\mathrm{ini}}$}}
\newcommand{\RcpRatio}{\mbox{$R_{\mathrm{CP}}^{R}/R_{\mathrm{CP}}^{\,0.2}$}}
\newcommand{\errordescr}{The error bars indicate statistical errors from the unfolding; the shaded boxes indicate point-to-point systematic errors that are only partially correlated. The solid lines indicate systematic errors that are fully correlated between all points.}

\begin{document}

\title{
Measurement of the jet radius and transverse momentum dependence of
inclusive jet suppression in lead-lead collisions
at $\sqrt{s_{\mathrm{NN}}}=2.76$~\TeV\ with the ATLAS detector}

\author{The ATLAS Collaboration}

\begin{abstract}
Measurements of inclusive jet suppression in heavy ion
collisions at the 
LHC provide direct sensitivity to the physics of jet quenching. 
In a sample of lead-lead collisions at $\sqrts = 2.76$~\TeV\ corresponding to an 
integrated luminosity of approximately $7~\mu \mathrm{b}^{-1}$, ATLAS has
measured jets with a calorimeter over the pseudorapidity interval
$|\eta| < 2.1$ and over the 
transverse momentum range $38 < \pT < 210$~\GeV. Jets were reconstructed using the
anti-\kt\ algorithm with values for the distance parameter that determines the 
nominal jet radius of $R = 0.2, 0.3, 0.4$ and
0.5. The centrality dependence of the jet yield is
characterized by the jet ``central-to-peripheral ratio,'' \Rcp. 
Jet production is found to be
suppressed by approximately a factor of two in the 10\% most central
collisions relative to peripheral collisions.  \Rcp\ varies
smoothly with centrality as characterized 
by the number of participating nucleons. The observed
suppression is only weakly dependent on jet radius and transverse
momentum. These results provide 
the first direct measurement of inclusive jet suppression in heavy ion
collisions and complement previous measurements of dijet transverse
energy imbalance at the LHC.
\end{abstract}
\maketitle

\section{Introduction}
Collisions of lead ions at the LHC are expected to create strongly
interacting matter at the highest temperatures ever produced in the
laboratory \cite{Abreu:2007kv}. This matter may be deconfined with a high density of
unscreened colour charges.
High transverse momentum (\pT) quarks and gluons generated by
hard-scattering processes have long been considered an important tool
for probing the properties of the matter created in ultra-relativistic
nuclear collisions. The energy loss of the partons propagating
through the matter may provide direct sensitivity to the colour
charge density and to the transport properties of the matter
\cite{Wang:1994fx,Baier:1998yf,Gyulassy:2000fs}. Indirect
observations of substantial parton energy loss or ``jet
quenching'' via suppressed single high-\pT\ hadron yields
\cite{Adler:2003qi,Adams:2003kv,Arsene:2003yk,Back:2004ra} and 
disappearance of the dijet contribution to di-hadron correlations
\cite{Adler:2002tq,phenix:2008cqb} have contributed to the conclusion 
that \AuAu\ collisions at RHIC produce a
quark-gluon 
plasma \cite{Adcox:2004mh,Adams:2005dq}. Observations of
highly asymmetric dijets in central \PbPb\ collisions at the LHC
\cite{Aad:2010bu, Chatrchyan:1327643,CMS:2012ni} can be understood in the context
of ``differential'' jet quenching, where one parton produced from an
initial hard-scattering loses significantly more energy than
the other, possibly as a result of different path lengths of the
partons in the matter \cite{Bjorken:1982tu}. However, the asymmetry is not
sensitive to situations where the two jets in a dijet pair lose
comparable amounts of energy, so other measurements are required to
probe ``inclusive'' jet quenching. 

The inclusive, per-event jet production rate provides such a
measurement. Energy loss of the parent partons in the created matter
may reduce or ``suppress'' the rate for producing jets at a given
transverse momentum. Such energy loss is expected to increase
with medium temperature and with increasing path length of the parton
in the medium \cite{Armesto:2011ht}. As a result, there should be more
suppression 
in central \PbPb\ collisions, which have nearly complete  
overlap between the incident nuclei, and little or no suppression in 
peripheral collisions where the nuclei barely overlap. In the absence
of energy loss, the jet production rate is expected to vary
with \PbPb\ collision centrality approximately in proportion to \Ncoll, the
number of nucleon-nucleon collisions that take place during a single
\PbPb\ collision. The jet suppression may be quantified
using the central-to-peripheral ratio, \Rcp, the 
ratio of the per-event jet yields divided by the number of
nucleon-nucleon collisions in a given centrality bin to the same
quantity in a peripheral centrality bin. The quantity, \Rcp, has the advantage
that potentially large systematic uncertainties, especially
those arising from systematic errors on the jet energy scale,
largely cancel when evaluating the ratios of jet spectra within the
same data set. The variation of the suppression with jet
transverse momentum and with collision centrality will depend both on
the energy loss mechanism and on the experimental definition of the
jet. In the case of radiative energy loss, jet energies can be reduced
by greater ``out-of-cone'' radiation, which should be more severe for
smaller jet radii \cite{Vitev:2008rz, Vitev:2009rd,He:2011pd}. 
Naively, collisional energy loss would result in a suppression that
is independent of radius. However recent calculations suggest that
collisional processes can also contribute to jet broadening \cite{Qin:2010mn}. A
measurement of the radius dependence of jet suppression could further clarify the
roles of radiative and collisional energy loss in jet quenching. 

This Letter presents measurements of the inclusive jet
\Rcp\ in \PbPb\ collisions at a nucleon-nucleon centre-of-mass
energy of $\sqrts = 2.76$~\TeV\ using data collected during 2010 
corresponding to an integrated luminosity of approximately $7~\mu
\mathrm{b}^{-1}$. Results are presented for jets reconstructed from 
energy deposits measured in the ATLAS calorimeters using the anti-\kt\
jet-finding algorithm \cite{Cacciari:2008gp}. The anti-\kt\ 
reconstruction was performed separately for four different values of
the anti-\kt\ distance parameter, $R$, that specifies the nominal radius of
the reconstructed jets, $R = 0.2, 0.3, 0.4$ and 0.5.  For the
remainder of the Letter the term ``radius'' will refer to the distance
parameter, $R$.
The jet energy is functionally defined to be the total energy within
the jet clustering algorithm above an uncorrelated underlying
event. This jet definition may include medium response with is
correlated with the jet. The underlying event contribution to each jet
was subtracted on a per-jet basis, and the
\Rcp\ values were calculated after unfolding the jet spectra for
distortions due to intrinsic jet resolution and underlying event fluctuations. 

\section{Experimental setup and trigger}
The measurements presented here were performed using the ATLAS
calorimeter, inner detector, trigger, and data acquisition systems
\cite{Aad:2008zzm}. The ATLAS calorimeter
system consists of a liquid argon (LAr) electromagnetic (EM) calorimeter
covering $|\eta|<3.2$, a steel-scintillator
sampling hadronic calorimeter covering $|\eta| < 1.7$, a LAr hadronic
calorimeter covering $1.5 < |\eta| < 3.2$, and two LAr electromagnetic and
hadronic forward calorimeters (FCal) covering $3.2 < |\eta| < 4.9$\footnote{ATLAS uses a
  right-handed coordinate system with its origin at the nominal
  interaction point (IP) in the centre of the detector and the
  $z$-axis along the beam pipe. The $x$-axis points from the IP to the
  centre of the LHC ring, and the $y$ axis points upward. Cylindrical
  coordinates $(r,\phi)$ are used in the transverse plane, $\phi$
  being the azimuthal angle around the beam pipe. The pseudorapidity
  is defined in terms of the polar angle $\theta$ as
  $\eta=-\ln\tan(\theta/2)$.} . The
hadronic calorimeter granularities or cell sizes in $\Delta \eta
\times \Delta \phi$ are $0.1 \times
0.1$ for $|\eta| < 2.5$ and  $0.2 \times 0.2$ for $2.5 < |\eta| < 4.9$\footnote{An 
  exception is the third (outermost) sampling layer, which has a segmentation of $0.2 \times 0.1$
up to $|\eta| = 1.7$.}.
The EM calorimeters are longitudinally segmented into three
compartments with an additional pre-sampler layer. The EM calorimeter 
has a granularity that varies with layer and pseudorapidity, but which
is generally much finer than that of the hadronic
calorimeter. The middle sampling layer, which typically has the
largest energy deposit in EM showers, has a $\Delta \eta \times \Delta \phi$ granularity of $0.025
\times 0.025$ over  $|\eta| < 2.5$. 

Charged particles associated with the calorimeter jets were measured
over the pseudorapidity interval $|\eta|<2.5$ using the
inner detector \cite{Aad:2010bx}. The inner detector is composed of
silicon pixel detectors in the innermost layers, followed by silicon
microstrip detectors and a straw-tube tracker, all immersed  
in a 2~T axial magnetic field provided by a solenoid.
Minimum bias \PbPb\ collisions were identified using measurements from
the zero-degree calorimeters (ZDCs) and the minimum-bias trigger
scintillator (MBTS) counters. The ZDCs are located symmetrically at $z = \pm
140$~m and cover $|\eta| > 8.3$. In \PbPb\ collisions the ZDCs
primarily measure ``spectator'' neutrons -- neutrons from the incident
nuclei that do not interact hadronically. The MBTS measures charged particles 
over $2.1 < |\eta| < 3.9$ using two sets of counters placed at $z = \pm 3.6$~m.
Events used in this analysis were selected for recording by the data
acquisition system using a logical or of ZDC and MBTS coincidence triggers.
The MBTS coincidence required at least one hit in each side of the
detector, and the ZDC coincidence trigger required the summed pulse
height from each calorimeter to be above a threshold set below the single neutron 
peak.

\section{Event selection and centrality definition}
In the offline analysis, \PbPb\ collisions were required to have a
primary vertex reconstructed from charged particle tracks with
\mbox{$\pT > 500$~\MeV}.  
The tracks were reconstructed from hits in the inner
detector using the standard ATLAS track reconstruction algorithm
\cite{Cornelissen:2008zzc} with settings optimized for the high hit
density in heavy ion collisions \cite{ATLAS:2011ah}. Additional
requirements of a ZDC coincidence, at least
one hit in each MBTS counter, and a difference in times measured by the
two sides of the MBTS detector of less than 3~ns were
imposed. The combination of the ZDC and MBTS 
conditions and the primary vertex requirement
efficiently eliminates both beam-gas interactions and 
photo-nuclear events ~\cite{Djuvsland:2010qs}. 
These event selections yielded a total of 51 million 
minimum-bias \PbPb\ events. Previous studies \cite{ATLAS:2011ah} 
indicate that the combination of trigger and offline
requirements select minimum-bias hadronic \PbPb\ collisions with an efficiency
of $98 \pm 2\%$. 

\begin{table}[b]
\caption{Results of Glauber model evaluation of $\langle \Npart
  \rangle$ and associated errors, $\langle \Ncoll \rangle$,
 the \Ncoll\ ratios, \Rcollcent,  and fractional errors on
 \Rcollcent\ for the centrality bins included in this analysis.  
}
\vspace{0.1in}
\centering
\begin{tabular}{| r | c | c | c | c | c  | } \hline
Centrality & $\langle \Npart \rangle$ & $\langle \Ncoll \rangle$ &
\Rcollcent \\ \hline
 0 -- 10\% &   $356 \pm 2$     &   $1500 \pm 115$ &   $57 \pm 6$    \\ \hline
10 -- 20\% &   $261 \pm 4$     &   $ 923 \pm 68$ &   $35 \pm 4$    \\ \hline
20 -- 30\% &   $186 \pm 4$     &   $ 559 \pm 41$ &   $21 \pm  2$    \\ \hline
30 -- 40\% &   $129 \pm 4$     &   $ 322 \pm 24$  &   $12 \pm  1$    \\ \hline
40 -- 50\% &   $ 86 \pm 4$ &   $ 173  \pm 14$ &   $6.5 \pm  0.04$    \\ \hline
50 -- 60\% &   $ 53 \pm 3$ &   $ 85 \pm 8$ &   $3.2 \pm  0.01$    \\ \hline
60 -- 80\% &   $ 23 \pm 2$ &   $ 27 \pm 4$ &   $1 $      \\ \hline
\end{tabular}
\label{tbl:glauber}
\end{table}
The centrality of \PbPb\ collisions was
characterized by \ETfcal, the total transverse energy measured in the
forward calorimeters. The distribution of \ETfcal\ was divided into
intervals corresponding to successive 10\% 
percentiles of the full centrality distribution after accounting for the missing
2\% most peripheral events. A standard Glauber Monte-Carlo analysis
\cite{Alver:2008aq,Miller:2007ri} was
used to estimate the average number of  participating
nucleons, $\langle \Npart \rangle$, and the average number of
nucleon-nucleon collisions, 
$\langle \Ncoll \rangle$, for \PbPb\ collisions in each of the centrality bins. 
The results are shown in Table~\ref{tbl:glauber}. The
\Rcp\ measurements presented here use the 60--80\% centrality bin as a
common peripheral reference. The \Rcp\ calculation requires the
ratio, $\Rcollcent \equiv \langle N_{\mathrm{coll}} \rangle
/\langle N_{\mathrm{coll}}^{\mathrm{\SixtyToEighty}} \rangle$, where $\langle
N_{\mathrm{coll}}^{\mathrm{\SixtyToEighty}} \rangle$ is the average number of
collisions in the 60--80\% centrality bin. The 
\Rcollcent\ uncertainties have been calculated by evaluating
the changes in \Rcollcent\ due to variations of the minimum-bias
trigger efficiency, parameters of the Glauber calculation, and
parameters in the modelling of the \ETfcal\ distribution
\cite{ATLAS:2011ah}. The \Rcollcent\ values and
uncertainties are also reported in 
Table~\ref{tbl:glauber}. 

\section{Monte Carlo samples}
Three Monte Carlo (MC) samples \cite{atlassim} were used for the
analysis in this Letter. A total of 1~million simulated minimum-bias
\PbPb\ events were produced using version 1.38b of the HIJING event generator
\cite{Wang:1991hta}. HIJING was run with default parameters except for
the disabling of jet quenching. To simulate the effects of elliptic
flow in \PbPb\ collisions, a parameterized centrality-, $\eta$- and
\pT-dependent $\cos{2\phi}$ modulation based on previous ATLAS
measurements \cite{ATLAS:2011ah} was imposed on the particles after
generation \cite{Masera:2009zz}. The detector response to the
resulting HIJING events was evaluated using GEANT4
\cite{Agostinelli:2002hh} configured with geometry and digitization
parameters matching those of the 2010 \PbPb\ run. 

A ``MC overlay'' data set, intended 
specifically for evaluating jet performance, was obtained by overlaying 
GEANT4-simulated $\sqrts = 2.76$~\TeV\ \pp\ hard-scattering events on the
HIJING events described above. The \pp\ events were obtained from the
ATLAS MC09 tune \cite{:2010ir} of the PYTHIA event generator
\cite{Sjostrand:2006za}. One million PYTHIA hard-scattering events
were generated for each of five intervals of \pthat, the transverse
momentum of outgoing partons in  the $2\rightarrow 2$ hard-scattering,
with boundaries $17, 35, 70, 140, 280$ and 560~\GeV. The \pp\ events
for each \pthat\ interval were overlaid on the same sample of HIJING events.

A smaller sample of ``data overlay'' events was produced by overlaying
150k GEANT4-simulated PYTHIA \pp\ events onto 150k 
minimum-bias \PbPb\ data events recorded during the 2011 LHC \PbPb\ run.
Due to the different detector conditions in the 2010 and 2011 runs,
the data overlay events cannot provide the corrections required for
this analysis. However, they provide a valuable test of the accuracy of
HIJING in describing the underlying event.

\section{Jet reconstruction}
\label{sec:recon}
Calorimeter jets were reconstructed from $\Delta \eta \times \Delta
\phi = 0.1\times 0.1$ towers using the anti-\kt\ algorithm \cite{Cacciari:2008gp} in
four-vector recombination mode with anti-\kt\ distance
parameters $R = 0.2, 0.3, 0.4$ and $0.5$. The tower energies were
obtained by summing energies, calibrated at the electromagnetic energy scale
\cite{Aad:2011he}, of all 
cells in all layers within the $\eta$ and $\phi$ boundaries of the
towers. Cells that span tower boundaries had their
energy apportioned by the fraction of the cell contained within a
given tower. The jet measurements presented here were obtained
by performing the anti-\kt\ reconstruction on the towers prior to
underlying event (UE) subtraction and then evaluating and subtracting
the UE from each jet at the calorimeter cell level. The subtraction
procedure calculates a per-event average UE energy density excluding
contributions from jets and accounting for effects of elliptic flow
modulation on the UE \cite{Poskanzer:1998yz}. The UE estimation and
subtraction was performed using a two-step procedure that was
identical for all jet radii. 

A first estimate of the UE average transverse energy 
density, ${\rho}_i(\eta)$, was evaluated 
in 0.1 intervals of $\eta$ from all cells in each calorimeter layer,
$i$, within  the given $\eta$ interval excluding those within
``seed'' jets. In the first subtraction step, the seeds are
defined to be $R = 0.2$ jets containing at least one tower
with $\ET > 3$~\GeV\ and having a ratio of maximum 
tower transverse energy to  average tower transverse energy, $
E_{\mathrm{T}}^{\mathrm{max}}/\langle E_{\mathrm{T}} \rangle > 
4$. Elliptic flow in \PbPb\ collisions
can impose a $2 v_2 \cos{\left[2(\phi - \Psi_2)\right]}$ modulation on the
UE. Here, $v_2$ is the second coefficient in a Fourier
  decomposition of the azimuthal variation of the UE particle or
  energy density, and the event plane angle, $\Psi_2$, determines the
  phase of the elliptic modulation.
Standard techniques \cite{ATLAS:2011ah,Poskanzer:1998yz} were
  used to measure $\Psi_2$, 
\begin{equation}
\Psi_2 = \frac{1}{2}\tan^{-1}{\left(\frac{\displaystyle \sum_{k} w_k
    {\ET}_k \sin{(2\phi_k)}}{\displaystyle \sum_k w_k {\ET}_k \cos{(2\phi_k)}}\right)},
\label{eq:psitwodef}
\end{equation}
where $k$ runs over cells in the FCal, $\phi_k$ represents the cell
azimuthal angle, and $w_k$ represent 
per-cell weights empirically determined to ensure a uniform $\Psi_2$
distribution. An $\eta$-averaged $v_2$ was
measured separately for each calorimeter layer according to
\begin{equation}
{{v_2}_i} = \frac{\displaystyle \sum_{j\in i} {\ET}_{j}
  \cos{\left[2\left(\phi_{j} -
      \Psi_2\right)\right]}}{\displaystyle \sum_{j\in i} {\ET}_{j}},
\label{eq:vtwodef}
\end{equation}
where $j$ runs over all cells in layer $i$. The UE-subtracted cell
transverse energies were calculated according to 
\begin{equation}
{E_{\mathrm{T}}}_j^{\mathrm{sub}} = {\ET}_j -
A_{j} \; {\rho}_i(\eta_j) \left (1 + 2\vtwo_{i} \cos{\left[2\left(\phi_j -
      \Psi_2\right)\right]}\right),
\label{eq:cellsub}
\end{equation}  
where ${\ET}_j$, $\eta_j$, $\phi_j$ and $A_j$ represent the cell 
\ET, $\eta$ and $\phi$ positions, and area, respectively
for cells, $j$, in layer $i$. 
The kinematics for $R = 0.2$ jets generated in this first subtraction step were
calculated via a four-vector sum of all (assumed massless) cells contained
within the jets using the \ET\ values obtained from Eq.~\ref{eq:cellsub}. 

The second subtraction step starts with the
definition of a new set of seeds using a combination of \RTwo\ jets
from the first subtraction step with $\ET > 25$~\GeV\ and track jets
(defined below) with $\pT > 10$~\GeV. Using this new set of seeds, a new estimate of
the UE, ${\rho}_i'(\eta)$, was calculated excluding cells within
$\Delta R = 0.4$ of the new seed jets, where $\Delta R =
\sqrt{(\eta_{\mathrm{cell}} - \eta_{\mathrm{jet}})^2 +
 (\phi_{\mathrm{cell}} - \phi_{\mathrm{jet}})^2}$.
New $\vtwo_i$ values, $\vtwo_i'$, were also calculated
excluding all cells within $\Delta \eta = 0.4$ of any of the new seed
jets. This exclusion largely eliminates distortions 
of the calorimeter $v_2$ measurement in events containing
high-\pt\ jets. The background subtraction was then applied to the 
original cell energies using Eq.~\ref{eq:cellsub} but with $\rho_i$
and $\vtwo_i$ replaced by the new values, ${\rho}_i'(\eta)$ and
$\vtwo_i'$. New jet kinematics were obtained for all jet
radii from a four-momentum sum of cells within the jets using the
subtracted cell transverse energies. Jets generated in this second
subtraction step having $\ET > 20$~\GeV\ were recorded for subsequent
analysis. 

A correction of typically a few per cent was applied to the reconstructed jets
to account for incomplete exclusion of towers within jets from the UE
estimate due, for example, to differences in direction between the seeds
and the final jets. This correction was validated by applying the full
heavy ion jet reconstruction procedure to 2.76~\TeV\ \pp\ data collected by
ATLAS in March 2011. The reconstructed jets were compared, jet-by-jet,
to those obtained from the \pp\ jet reconstruction procedure. After
this last correction for incomplete exclusion of jets from the
background, the energy scales of the heavy ion and \pp\ reconstruction
procedures agreed to better than $1\%$ for $\ET > 25$~GeV. A final 
correction depending on the jet $\eta$, \ET, and $R$ 
was applied to obtain the correct hadronic energy scale for the
reconstructed jets. The calibration constants were derived separately
for the four jet radii using the same procedure applied to \pp\ jet measurements \cite{Aad:2011he}.

In addition to the calorimeter jet reconstruction, track
jets were reconstructed using the anti-\kt\ algorithm with 
$R = 0.4$ from charged tracks that have a good match to the
primary vertex and that have $\pT > 4$~\GeV. This
threshold suppresses contributions of the UE to  
the track jet measurement. Specifically, an $R = 0.4$ track jet 
has an estimated likelihood of including an uncorrelated $\pt > 4$~\GeV\
charged track of less than $4\%$ in the 0--10\% centrality bin. The
single track reconstruction efficiency is $\approx80\%$, approximately independent of centrality.

\begin{figure*}[t]
\centerline{
\includegraphics*[height=3.25in]{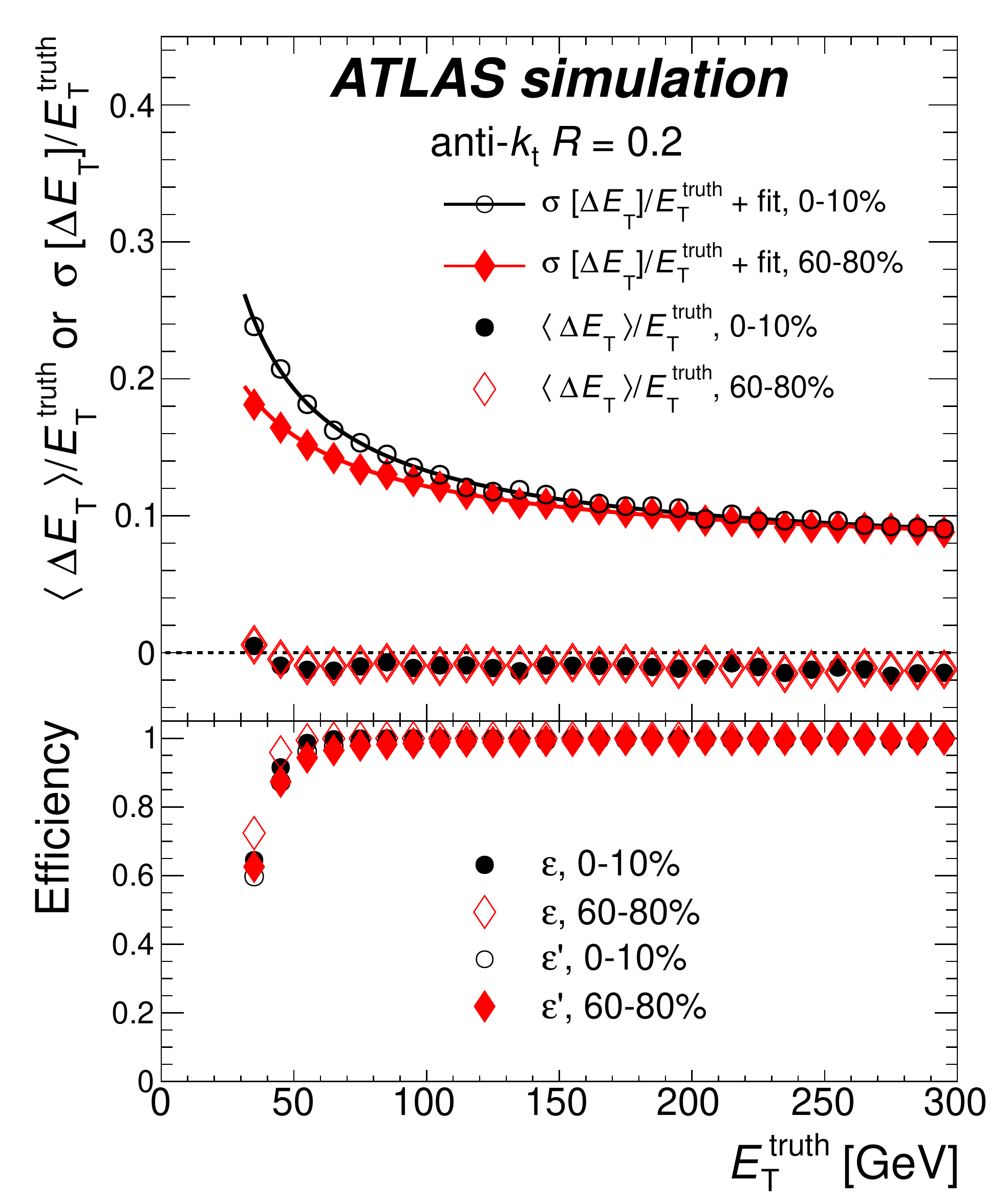}
\hspace{-0.15in}
\includegraphics*[height=3.25in]{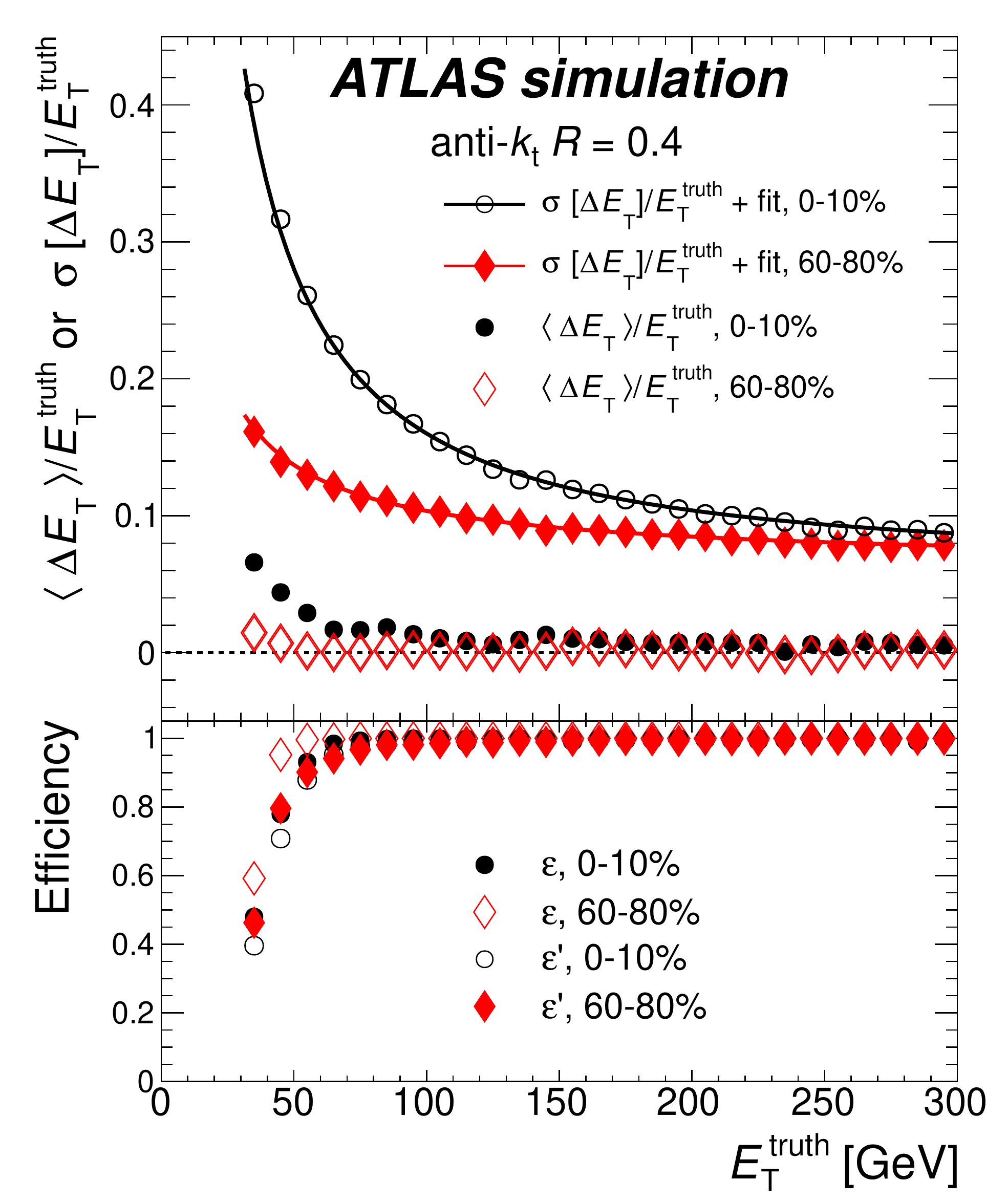}
}
\caption{Results of MC evaluation of jet reconstruction performance in 0--10\% and
  60--80\% collisions as a function of truth jet \ET\ for
  \RTwo\ (left) and \RFour\ (right) jets. Top: jet energy resolution
  $\sigma[{\DEt}]/\ETtrue$ and jet energy scale closure, $\langle
  \DEt\rangle/\ETtrue$. Solid curves show parameterizations of the JER 
using Eq.~\ref{eq:resolparam}.
Bottom: Efficiencies, $\varepsilon$ 
and $\varepsilon'$, for reconstructing jets 
  before and after application of UE jet removal (see text for
  explanation), respectively. 
}
\label{fig:perform}
\end{figure*}
The fluctuating UE in \PbPb\ collisions can potentially produce
reconstructed jets that do not originate from hard-scattering
processes. In the remainder of this Letter such jets are referred
to as ``underlying event jets'' or UE jets. A requirement that
calorimeter jets match at least one track jet with $\pt > 7$~\GeV\ or
an EM cluster reconstructed from cells in the electromagnetic 
calorimeter \cite{Aad:2009wy} with $\pt > 7$~\GeV\ was applied to
exclude UE jets. The matching criterion for
both track jets and EM clusters 
is that they lie within $\Delta R = 0.2$ of the jet. 
Applying this matching requirement provides a factor of about 50 rejection 
against UE jets while inducing an additional \pT-dependent
inefficiency in the jet measurement. To accommodate the use of track
jets in the UE jet rejection, the jet measurements presented here
have been restricted to $|\eta| < 2.1$. The total number of jets above
\pt\ thresholds of 40~\GeV\ and 100~\GeV\ in the data sample after event
selection, UE jet rejection, and the $|\eta| < 2.1$ cut have been applied
is shown in Table~\ref{tbl:jet_stats} for the most central and
peripheral bins.
\begin{table}
\caption{Total number of jets in the data set with $\pt > 40$~\GeV\ and $\pt > 100$~\GeV\ in the
  0--10\% and 60--80\% centrality bins after all event selection
  criteria, UE jet rejection, and the $|\eta| < 2.1$ cut have been
  applied.}
\vspace{0.1in}
\centering
\begin{tabular}{|c|r|r|r|r|}\hline
&  \multicolumn{2}{c|}{$\pt>40$~\GeV} &  \multicolumn{2}{|c|}{$\pt>100$~\GeV}
\\  \cline{2-5}
 $R$ &0--10\%& 60--80\%& 0--10\%&60--80\% \\ \hline
             0.2 &         112 333 &          8068 &  2308 &                      162\\ \hline
              0.3 &         287 153 &         12 629 &  3534 &                      222\\ \hline
              0.4 &         543 444 &         15 964 &  4974 &                      277\\ \hline
              0.5 &         710 158 &         18 573 &  7586 &                      307\\ \hline
\end{tabular}
\label{tbl:jet_stats}
\end{table}
\section{Performance of the jet reconstruction}
\label{sec:perform}
The primary evaluation of the combined performance of the ATLAS
detector and the analysis procedures described above in measuring
unquenched jets was obtained using the MC overlay sample. In that MC
sample, the kinematics of the reference PYTHIA generator-level jets
(hereafter called ``truth jets'') were reconstructed from PYTHIA
final-state particles for $R = 0.2, 0.3, 0.4$ and $0.5$ using the
same techniques as applied in \pp\ analyses
\cite{Aad:2011he}. Separately, the presence and approximate kinematics
of HIJING-generated jets were obtained by running $R = 0.4$
anti-\kt\ reconstruction on final-state HIJING particles having $\pT
>4$~\GeV. Accidental overlap of jets from unrelated hard-scattering
processes may occur at non-negligible rates in the data due to
the geometric enhancement of hard-scattering rates in \PbPb\
collisions. However, for the purposes of this Letter, the resulting
combined jets are considered part of the physical jet spectrum and
not a result of UE fluctuations. Then, to prevent the overlap of PYTHIA
and HIJING jets from distorting the jet performance evaluated relative
to PYTHIA truth jets, all PYTHIA truth jets within $\Delta R = 0.8$ of
a $\pT > 10$~\GeV\ HIJING jet were excluded from the analysis.  

Following reconstruction of the overlaid MC events using the same
algorithms that were applied to the data, PYTHIA truth jets passing
the HIJING-jet exclusion were matched to the closest reconstructed jet
of the same $R$ value within $\Delta R = 0.2$. The resulting matched
jets were used to evaluate the jet energy resolution (JER) and the jet
energy scale (JES). The jet reconstruction efficiency was defined as
the fraction of truth jets for which a matching reconstructed jet is
found. The efficiency was evaluated both prior to ($\varepsilon$) and
following ($\varepsilon'$) UE jet rejection. For all three performance
measurements, the different \phat\ MC overlay samples were combined
using a weighting based on the PYTHIA cross-sections for each
\pthat\ range. 

Figure~\ref{fig:perform} shows a summary of the ATLAS 
\PbPb\ jet reconstruction performance for \RTwo\ and \RFour\ jets in central (0--10\%)
and peripheral (60--80\%) collisions. The (fractional) JER was
characterized by $\sigma[\DEt]/\ETtrue$, where $\sigma[\DEt]$ is the
standard deviation of the $\DEt \equiv \ETrec - \ETtrue$ distribution
and where \ETrec\ and \ETtrue\ are the 
reconstructed and truth jet \ET\ values, respectively. The JES offset
or ``closure''  was evaluated from the mean fractional energy shift,
$\langle \DEt\rangle/\ETtrue $.
\begin{figure*}
\centerline{
\includegraphics[height=3.8in]{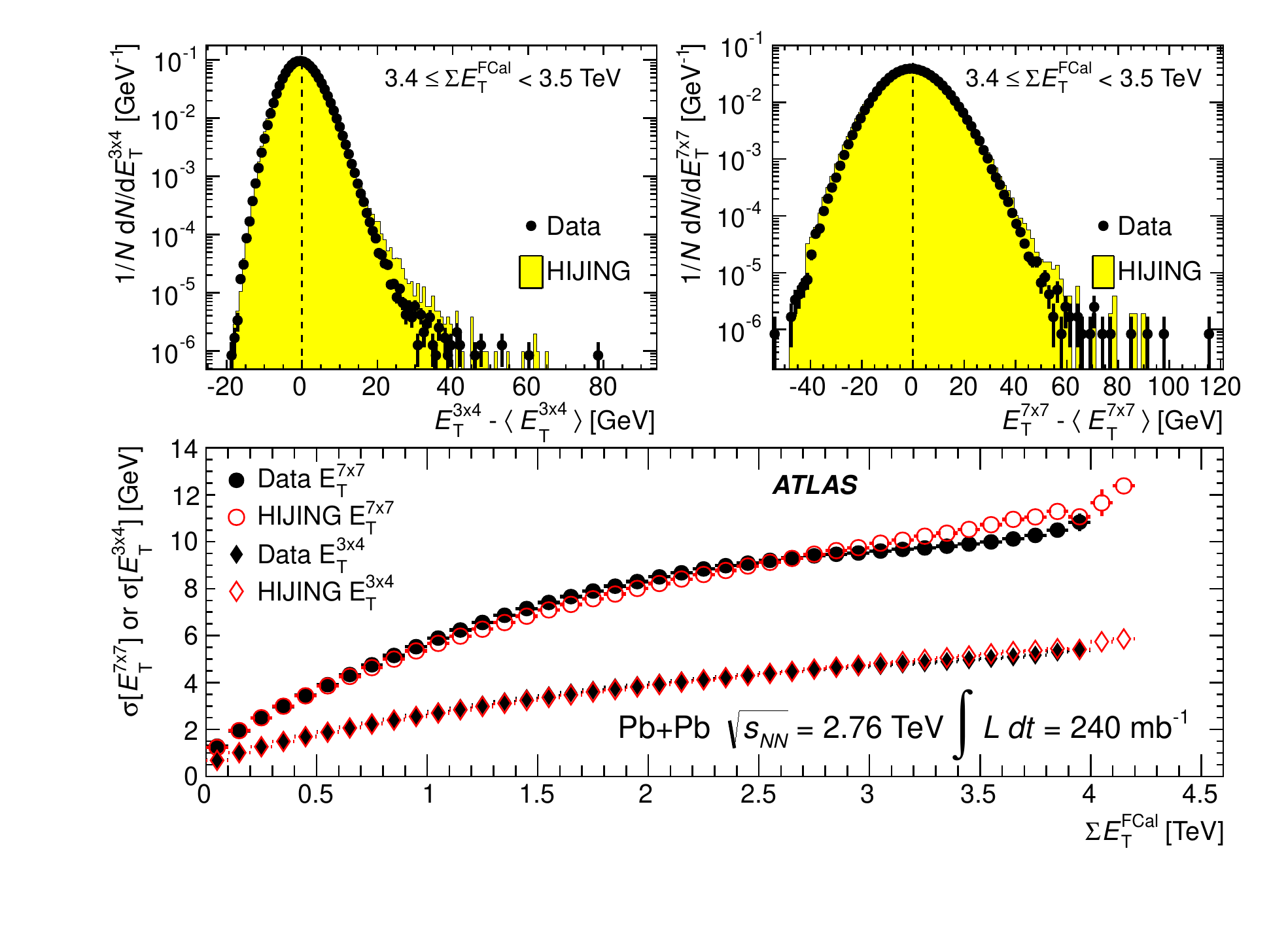}
}
\vspace{-0.4in}
\caption{Top: Representative distributions of $\ETtbyf - \langle\ETtbyf\rangle$
  (left) and $\ETsbys - \langle\ETsbys\rangle$ (right)  (see text for
  definitions) for data (points) and MC (filled histogram) for
  \PbPb\ collisions with $3.4 \leq \ETfcal < 3.5$~\TeV. The vertical lines
  indicate $\ETtbyf - \langle\ETtbyf\rangle = 0$ and $\ETsbys -
  \langle\ETsbys\rangle = 0$.
  Bottom: Standard deviations of the \ETtbyf\ and
  \ETsbys\ distributions, $\sigma[\ETtbyf]$ and $\sigma[\ETsbys]$,
  respectively, in data and HIJING MC sample as a function of \ETfcal.   
}
\label{fig:fluct}
\end{figure*}

The JER was found to be well described by a quadrature sum of three
terms, 
\begin{equation}
\frac{\sigma[\DEt]}{\ETtrue} = \frac{a}{\sqrt{\ETtrue}} \oplus \frac{b}{\ETtrue} \oplus c,
\label{eq:resolparam}
\end{equation}
where $a$ and $c$ represent the
usual sampling and constant contributions to calorimeter resolution.
The term containing $b$ describes the contribution of underlying event
fluctuations, which do not depend on jet \ET. Results of
fitting the \ET\ dependence of the JER according to
Eq.~\ref{eq:resolparam}, using methods described below, are shown with
curves in Fig.~\ref{fig:perform}. 

The jet reconstruction efficiency decreases with decreasing jet
\ET\ for $\ET \lesssim 50$~\GeV. The decrease starts at larger \ET\ and
decreases more rapidly for larger jet radii and in more central
collisions. The inefficiency results primarily from the finite JER which
causes jets with $\ETtrue > 20$~\GeV to be measured with $\ETrec <
20$~\GeV. The UE jet rejection causes an additional loss of jets but in
a manner that reduces the centrality dependence of the inefficiency. 

The accuracy of the MC overlay studies described above was evaluated
using the data overlay sample analyzed using the same procedures
that were applied to the MC overlay sample. The analysis yielded
results for the JER, JES, and efficiency consistent with the
MC overlay sample, although the JER in the data overlay sample
was found to be slightly better than in the MC overlay sample.
The JES in the data overlay sample was found to agree between
peripheral and central collisions to better than 1\% for
\RFour\ jets, and the reconstruction efficiency was found to differ
 by less than 5\% on the rise of the efficiency curve.

A data-driven check of the HIJING description of UE
fluctuations was performed by evaluating distributions of
EM-scale summed \ET\ in rectangular groups of
towers within the interval $|\eta| < 2.8$. The groups were chosen to
match the areas of jets used in this analysis: $3\times 4$ and $7
\times 7$ for \RTwo\ and \RFour\ jets, respectively. No attempt was
made to exclude jets from the fluctuation analysis. The distributions of \ETtbyf\ and \ETsbys, the $\Sigma \ET$ for
$3\times 4$ and $7\times 7$ groups of towers, are shown in
Fig.~\ref{fig:fluct} for a narrow range of \ETfcal, $3.4 \leq \ETfcal <
3.5$~\TeV, that lies within the 0--1\% centrality interval.
These distributions have mean values, 
$\langle\ETtbyf\rangle=26$~\GeV\ and $\langle\ETsbys\rangle=105$~\GeV,
subtracted and, thus, in principle represent
the distribution of the residual contributions of the UE to jet energies
after subtraction. However, the high tails of the distributions can be
attributed to the presence of jets, which are not part of the UE. The corresponding distributions obtained
from the HIJING MC sample, but with $\langle\ETtbyf\rangle$ and
$\langle\ETsbys\rangle$ obtained from data, are shown in
Fig.~\ref{fig:fluct} with filled histograms. 

The shapes of the MC and data distributions in Fig.~\ref{fig:fluct}
 (top) are very similar, but the MC result slightly over-predicts the 
positive fluctuations for all collision centralities. In central
collisions the MC result also slightly over-predicts the size of negative 
fluctuations. In contrast, for non-central collisions (not shown here)
the data has a broader distribution of negative 
fluctuations than the MC sample. These observations are demonstrated by
Fig.~\ref{fig:fluct} (bottom) which shows the standard
deviations of the \ETtbyf\ and \ETsbys\ distributions,
$\sigma[\ETtbyf]$ and $\sigma[\ETsbys]$, as a function of \ETfcal,
obtained from both the data and the MC sample. The data and MC
distributions have similar trends, but the MC
$\sigma[\ETtbyf]$ and $\sigma[\ETsbys]$ values are larger in central
collisions by 2.5\% and 5\%, respectively. In non-central collisions,
the broader spectrum of negative fluctuations in the data causes
$\sigma[\ETtbyf]$ and $\sigma[\ETsbys]$ to exceed the corresponding
quantity in the HIJING MC sample by approximately the same percentages. 

Consistency between the results of the fluctuation analysis and the
evaluation of the JER described above has been established by fitting
the \ET\ dependence of the JER with the functional form
given by Eq.~\ref{eq:resolparam} with fixed $b$ values obtained from the
fluctuation analysis. The $b$ values for a given jet radius were
determined by taking the standard deviation of the $\Sigma \ET$
distribution for the corresponding tower group averaged over centrality
and corrected to the hadronic energy scale. The resulting $b$ values
for \RTwo($0.4$) jets are $5.62(12.45)$~\GeV\ and $1.15(2.58)$~\GeV\ for
the 0--10\% and 60--80\% centrality bins respectively. The parameters
$a$ and $c$ obtained from the fits are found to be independent of
centrality within fit uncertainties, as expected, and to have values $a = 1.0
(0.8)$, $c = 0.07 (0.06)$ for \RTwo($0.4$) jets with \ET\ expressed in
\GeV. The accuracy of the fits in describing the \ET\ dependence of
the JER is demonstrated by the curves showing results for \RTwo\ and
\RFour\ jets in Fig.~\ref{fig:perform}.

The contribution of UE jets to the measured jet spectrum after UE jet
rejection is estimated to be $\lesssim 3\%$ approximately independent
of jet \pT\ for $40 < \pT < 60$~\GeV\ and less than 1\% for $\pT >
60$~\GeV. This estimate was obtained by evaluating the rate of
reconstructed jets in the HIJING MC sample which were not matched to 
HIJING truth jets and correcting for missing truth jets due to the
$\pT > 4$~\GeV\ requirement applied in the HIJING truth jet
reconstruction. 

\section{Jet spectra and unfolding}
\begin{figure}
\centerline{
\includegraphics[width=0.48\textwidth]{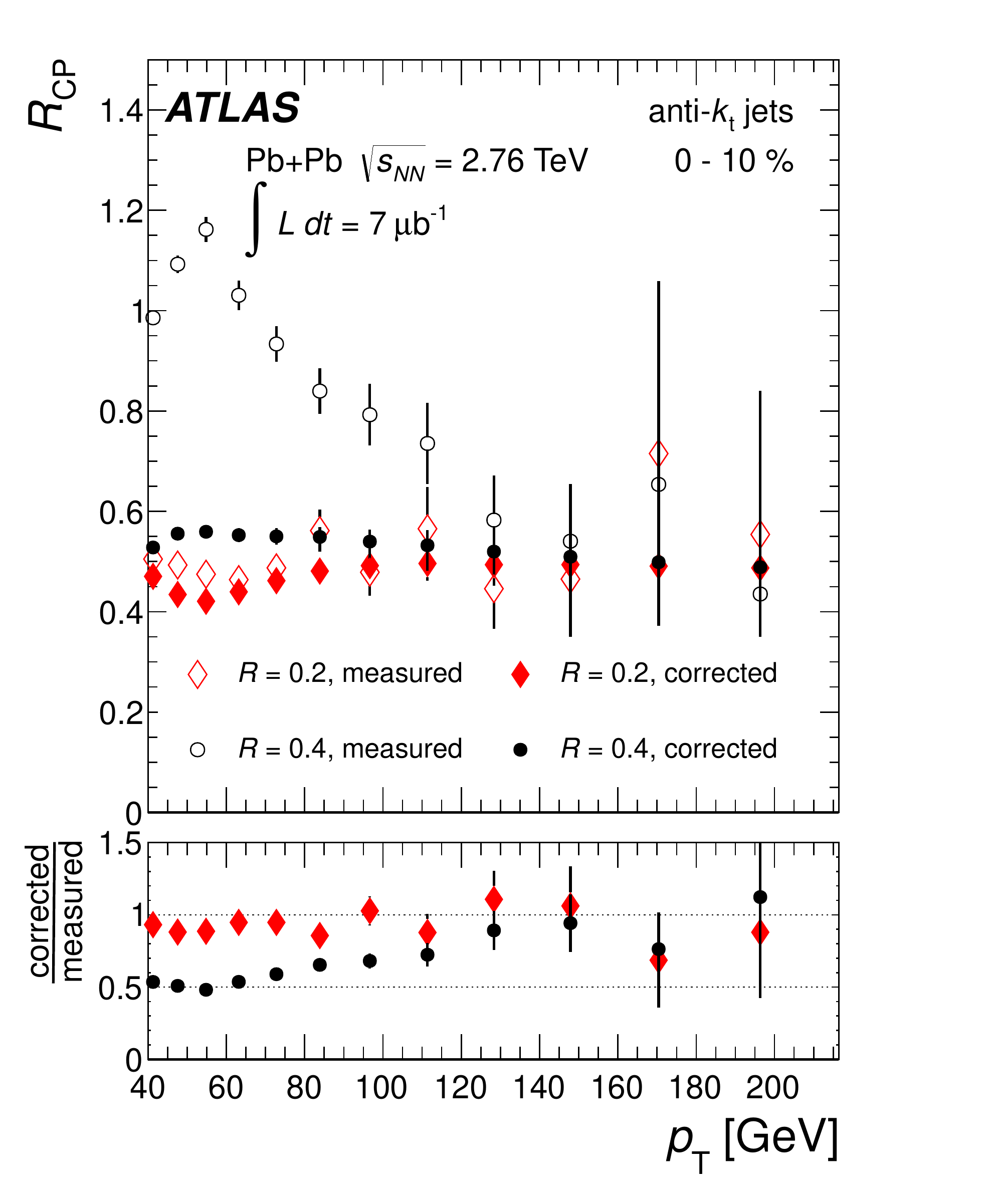}
}
\caption{
Top: Measured and corrected \Rcp\ values for
the 0--10\%  centrality bin as a function of jet
\pT\ for \RFour\ and \RTwo\ jets. Bottom: Ratio of 
corrected to measured \Rcp\ values for both
jet radii. The error bars on the points represent statistical
uncertainties only. 
}
\label{fig:corruncorrrcp}
\end{figure}

Though jet reconstruction performance is naturally evaluated in terms
of jet \ET, the 
physics measurements in this Letter were performed as a function of
\pt\ directly calculated from the jet four-momentum. The typical
masses of the jets are sufficiently small 
that $\ET \approx \pt$ holds over the range of measured \pt\ for all
jet radii. The measured \pt\ spectra of  
reconstructed jets passing UE jet rejection and having $|\eta| < 2.1$
were evaluated for each centrality bin using logarithmic \pT\ bins
spanning the range $38 < \pT < 210$~\GeV. The correlations within and
between \pT\ bins arising from multi-jet events were quantified by the
covariance, $C_{ij}$, between the number of jets measured in two bins,
$i$ and $j$. The measured \Rcp\ was calculated as 
\begin{equation}
\left.\rcpraw(\pT)\right|_{\mathrm{cent}} =
\dfrac{1}{ R_{\mathrm{coll}}\centup} \left(
\dfrac{
\dfrac{ N_{\mathrm{jet}}\centup(\pT)}{ N_{\mathrm{evt}}\centup}
}{
 \dfrac{  N^{\SixtyToEighty}_{\mathrm{jet}}(\pT)}{ N_{\mathrm{evt}}^{\SixtyToEighty}}
}\right),
\label{eq:rcprawcalc}
\end{equation}
where $N_{\mathrm{jet}}\centup$ represents the measured jet yield in a given
\pT\ and centrality bin, and
$N_{\mathrm{evt}}\centup$ and $N_{\mathrm{evt}}^{\SixtyToEighty}$ are the number of
\PbPb\ collisions within the
chosen and peripheral reference centrality intervals, respectively. Results for
$\rcpraw|_{\ZeroToTen}$ obtained from the measured spectra are shown in
Fig.~\ref{fig:corruncorrrcp} for \RTwo\ and \RFour\ jets. The
$\rcpraw|_{\ZeroToTen}$ for \RTwo\ jets is approximately equal to 0.5 over
the measured \pT\ range. 
The $\rcpraw|_{\ZeroToTen}$ for \RFour\ and \RTwo\ jets are consistent for
$\pt > 120$~\GeV, but at lower \pt, the \RFour\ $\rcpraw|_{\ZeroToTen}$ increases
relative to the \RTwo\ values. The difference between \RTwo\ and
\RFour\ $\rcpraw|_{\ZeroToTen}$ values can be mostly attributed to
the difference in the size of the UE fluctuations for \RTwo\ and
\RFour\ jets shown in Fig.~\ref{fig:perform}. The larger JER for
\RFour\ jets produces greater upward migration on the steeply falling
jet \pT\ spectrum in central collisions than in peripheral collisions,
thus enhancing the measured \Rcp. The drop in the
\RFour\ $\rcpraw|_{\ZeroToTen}$ 
at low \pT\ is due to the decrease in jet reconstruction
efficiency between 60-80\% and 0-10\% centrality bins which, as noted
above, largely results from the worse JER in central collisions.

To remove the effects of the bin migration, the jet spectra were
unfolded using the singular value decomposition (SVD) technique
\cite{Hocker:1995kb} as implemented in
\verb+RooUnfold+\ \cite{Adye:2011gm}. The 
MC overlay samples were used to populate a response matrix, {\boldmath{$A$}, which
describes the transformation of the true jet spectrum, \boldmath{$x$},  
to the observed spectrum, \boldmath{$b$}, according to \boldmath{$b =
  Ax$}\unboldmath. The truth and reconstructed jet \pt\ were obtained
from the MC overlay sample using the methods described in
Section~\ref{sec:perform}, and \ref{sec:recon}, respectively, and the
selection and matching of truth and reconstructed jet pairs was performed as
described in Section~\ref{sec:perform}.
Using the weighting method suggested in Ref.~\cite{Hocker:1995kb}, the
unfolded spectrum is expressed as a set of weights {\boldmath $w$} multiplying 
the input spectrum (\xini) used to produce {\boldmath{$A$}}. 
The SVD method expresses the solution for {\boldmath $w$}  
in terms of a least-square minimization problem that includes a
prescription for regularizing the amplification of statistical
fluctuations of the data that would result from the direct inversion
of \boldmath{$A$}}\unboldmath. The regularization is controlled by a parameter
$\tau$ such that contributions from singular values $s_k$ of the
unfolding matrix with $s_k < \tau$ are suppressed. 
Inclusion of the \pT-dependent reconstruction efficiency in the
response was found to strongly affect the spectrum of singular values
of the matrix defining the SVD problem, so the efficiency correction
was applied separately following the unfolding. The spectrum of MC
truth jets was reweighted to provide a smooth, power-law initial spectrum, $\xini
\propto \varepsilon'(\pT)/p_{\mathrm{T}}^{n}$, where the power index
was chosen to be $n = 5$. An analysis of the 
optimal regularization in the SVD unfolding following the methods of
Ref.~\cite{Hocker:1995kb} indicated that a regularization parameter fixed
by the fifth singular value ($\tau = s_5^2$) of the SVD matrix was
appropriate for all centralities and all $R$ values. The statistical
uncertainties in the SVD unfolding due to statistical errors on the
input spectrum were evaluated using the pseudo-experiment technique
with 1000 separate stochastic variations of the input spectrum based
on the full covariance matrix. The contributions of statistical
fluctuations in the response matrix, {\boldmath $A$}, were similarly
evaluated using an equal number of stochastic variations of the
response matrix. The two contributions to the statistical uncertainty  were
combined in quadrature. 

Potential biases in the unfolding procedure were evaluated using two
different methods. Each unfolded spectrum was re-folded with its
corresponding response matrix and compared to the measured spectrum for
self-consistency. In general, regularization can introduce differences
between re-folded and measured spectra on the scale of statistical
uncertainties on the measured spectra, while over-regularization 
can produce larger, systematic differences. For all of the unfolded
spectra, the re-folding procedure yielded a typical difference between
measured and re-folded spectra comparable to the statistical
uncertainties on the measured spectra. A separate check
was performed by unfolding the reconstructed MC spectrum for each
centrality bin and each jet radius and comparing to the original MC
truth jet spectrum. For this purpose, the MC data sets were divided 
in half and reconstructed spectra and response matrices were generated
separately from each set. The unfolded and truth MC jet spectra
typically agreed to better than 2\%, though for the 0--10\% centrality
bin and for $R = 0.4$
and 0.5 jets, differences as large as 5\% were observed in the lowest
\pt\ bins. These
differences are covered by the unfolding systematic uncertainties described below.

The corrected \Rcp\ was evaluated according to
\begin{equation}
\left.\rcpcorr(\pT)\right|_{\mathrm{cent}} = \frac{1}{R_{\mathrm{coll}}\centup}\left(
\dfrac{\dfrac{{\tilde{N}_{\mathrm{jet}}}\centup(\pT) }{{\varepsilon'}\centup \;
  N_{\mathrm{evt}}\centup}}{
\dfrac{ \tilde{N}_{\mathrm{jet}}^{\SixtyToEighty}(\pt)}{\varepsilon'^{\SixtyToEighty} \; N_{\mathrm{evt}}^{\SixtyToEighty}}}\right),
\label{eq:rcpunfcalc}
\end{equation}
where $\tilde{N}_{\mathrm{jet}}$ represents the unfolded number of jets in the \pT\ bin, and  
${\varepsilon'}\centup$ and ${\varepsilon'}^{\SixtyToEighty}$ are the
  \pt-dependent jet reconstruction efficiencies after UE jet
  rejection for the indicated centrality bins. Figure~\ref{fig:corruncorrrcp} shows the 
comparison of the corrected and measured \Rcp\ values as a function
of jet \pT\ for \RTwo\ and \RFour\ jets in the 0--10\% centrality bin.
The unfolding has little effect on the \RTwo\ \Rcp\ due to the good
energy resolution (relative to larger radii) for \RTwo\ jets even in
central collisions. For the 
\RFour\ jets, \Rcp\  is reduced by a factor of about two at the 
lowest \pt\ values included in the analysis and is only slightly
modified at the highest \pT. Because the unfolding provides a non-local mapping of the
input jet \pT\ spectrum onto the unfolded spectrum, the statistical
uncertainties in the unfolded spectra have significant correlations
between bins, and there is not a direct relationship between the
statistical errors in the input spectrum and the unfolded
spectrum. The regularization of the unfolding also suppresses
statistical fluctuations in the unfolded spectrum, but the statistical
uncertainties in the measured spectrum also contributes to the systematic
uncertainties from the unfolding procedure. 

\section{Systematic uncertainties}
Systematic uncertainties in the \Rcp\ measurement can arise due to
errors on the jet energy scale (JES), the jet energy
resolution (JER), jet finding efficiency, the unfolding
procedure, and the \Rcollcent\ values. Uncertainties in jet \ET\ and
\pt\ are assumed to be equal (i.e. $\delta \pt = \delta
\ET$). Uncertainties in the JES and the JER influence the unfolding of
the jet spectra. The resulting systematic uncertainties on the
\Rcp\ values (\delRcp) were evaluated by producing new response
matrices according to the procedures described below,  generating 
unfolded spectra from these matrices, and calculating new
\Rcp\ values. The resulting changes in the \Rcp\ values were taken to
be estimates of \delRcp. For uncertainties fully correlated in
centrality, \delRcp\ was evaluated by simultaneously varying the
chosen centrality bin and the 60--80\% bin, while for other
uncertainties, the chosen centrality bin and 60--80\% centrality bins
were varied separately and the variations in \Rcp\ combined in
quadrature.

Overall JES uncertainties
common to the different centrality bins cancel in the ratio of the
spectra in \Rcp, but centrality-dependent JES errors will produce
a systematic shift in \Rcp. Studies using the MC overlay sample
discussed in Section~\ref{sec:perform} indicate a maximum
difference in JES between the 0--10\% and 60--80\% centrality
bins for the jet \pt\ range included in this analysis of 0.5\%, 1\%,
1.5\% and 2.5\% for $R = 0.2, 0.3, 0.4$ and $0.5$, respectively. 
Studies were also performed with the data overlay sample
using an identical procedure as that applied to the MC overlay
sample. The JES evaluated in the data overlay was found to agree between the
0--10\% and 60--80\% centrality bins to better than 1\%, which is better
than the agreement found in the MC overlay sample.

Independent evaluations of a possible centrality dependence of the
calorimeter JES were performed by matching track and calorimeter jets
in both the data and the MC overlay sample. The track jets provide a
common reference for evaluating calorimeter jet response 
that is insensitive to the UE. The average calorimeter
jet \ET\ was evaluated as a function of matching track jet \pt, $\langle E_{\mathrm{T}}^{\mathrm{calo}}\rangle 
(\pttrkjet)$, for different centrality
bins. In the data, for $\pttrkjet > 50$~\GeV, the $\langle
E_{\mathrm{T}}^{\mathrm{calo}}\rangle$ values were found to be 
consistent across all centrality bins to better than 3\%. Accounting for a slight
centrality dependence seen in the MC overlay sample, the 0--10\% and 60--80\% bins agree
to 2\%. For  $\pttrkjet < 50$~\GeV, $R$- and centrality-dependent differences of up
to 4\% (for \RFive) are observed between data and MC overlay results
for $\langle E_{\mathrm{T}}^{\mathrm{calo}}\rangle
(\pttrkjet)$. This study provides a stringent
constraint on changes in calorimeter response for jets affected by
quenching and justifies the use of unquenched jets from PYTHIA in evaluating the
jet performance and response matrices.

\begin{figure}[t]
\centerline{
\includegraphics*[width=0.47\textwidth]{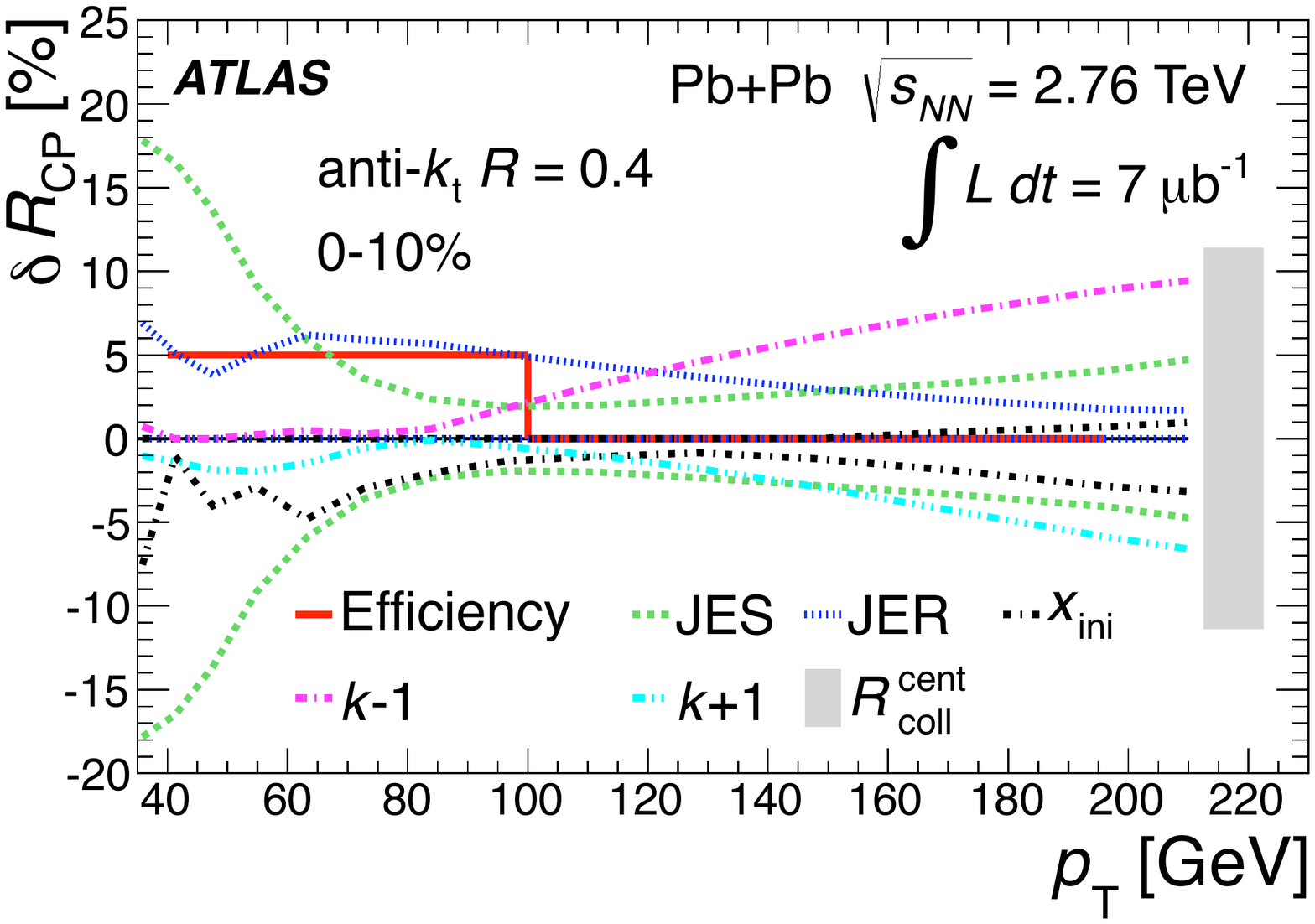}
}
\caption{Contributions to the relative systematic uncertainty on the
  \Rcp\ from
  various sources for the \RFour\ \antikt\ jets in the 0--10\%
  centrality bin. The $k\pm1$ curves denote the uncertainty due to the
  choice of regularization parameter obtained by unfolding with
the fourth and sixth singular values. A constant 5\% systematic uncertainty
on the jet reconstruction efficiency is assigned for $\pT < 100$~\GeV only.
The 11\% uncertainty in the determination
  of \Rcollcent\ is indicated with a shaded box and is
  \pt-independent. }
\label{fig:syserror}
\end{figure}
Based on the combination of the studies described above, the systematic uncertainties
on the centrality dependence of the JES for the 0--10\%
centrality bin and for calorimeter jet $\pt > 70$~\GeV\ were estimated
to be 0.5\%, 1\%, 1.5\% and 2.5\%, respectively, for $R = 0.2, 0.3, 0.4$ and $0.5$
jets. At lower \pt, the assigned systematic uncertainties 
increase linearly with decreasing \pt\ such that they double in size
between 70~\GeV\ and 38~\GeV. For other centrality bins, the
systematic errors on the centrality dependence of the JES decrease
smoothly from central to peripheral collisions.
The resulting \delRcp\ values were evaluated using 
new response matrices generated by scaling the reconstructed \pt\ to
account for the above-quoted JES uncertainties. The JES systematic uncertainty is
assumed to be fully correlated between different centrality bins and
different $R$ values.

Systematic uncertainties in the JER due to inaccuracies in the MC
description of the UE fluctuations were evaluated using results of
the fluctuation analysis described above. The effects of those
inaccuracies were evaluated by rescaling the per-jet $\Dpt \equiv
\pTrec - \pTtrue$ values obtained from the MC study by 
factors that cover the differences between data and MC result.
For each centrality and jet radius, a modified value of the $b$
parameter in Eq.~\ref{eq:resolparam} was evaluated 
and used to obtain new JER values, $\sigma'[\DEt]$ 
from Eq.~\ref{eq:resolparam}. Then a rescaled \Dpt\ was obtained from 
\begin{equation}
\Dpt' = \Dpt \left(
\frac{\sigma'}{\sigma}\right).
\end{equation}
Since the discrepancies between the MC and the data were observed to be
different for positive and negative fluctuations, the rescaling was
applied separately for positive and negative \Dpt. 

The $\Sigma
\ET$ values in the MC study were found to have larger positive fluctuations 
than those in the data for all centralities by approximately 2.5\%,
2.5\%, 5\%, and 7.5\% for $R = 0.2, 0.3, 0.4$ and 0.5 jets,
respectively, so for positive \Dpt\ $b$ was reduced by these
percentages. For the 0--10\% centrality bin, the negative fluctuations were
also larger in the MC study than in the data by the same approximate
percentages, so for central collisions the same, modified $b$ value was 
used for negative \Dpt. For all other centrality bins, the negative
fluctuations in the data were larger than in the MC by approximately
twice the above-quoted percentages. Thus, for those centralities, the
modified $b$ values were obtained for negative \Dpt\ by increasing $b$
by 5\%, 5\%, 10\%, and 15\%, respectively, for $R = 0.2, 0.3, 0.4$ and
0.5 jets. 

New response matrices were generated using the calculated
$\Dpt'$ values according to $\pTrec' = \pTtrue + \Dpt'$, and these
modified response matrices were used to 
estimate the JER systematic uncertainties following the procedure
described above. The systematic uncertainty on the spectra due to the JER 
for the 0-10\% centrality bin was taken to be one-sided as all
evaluations indicate that the MC simulations slightly overestimate UE
fluctuations. Asymmetric errors were obtained for
the other centrality bins by applying
the positive and negative \DEt\ scalings separately. The JER
systematic uncertainties were assumed to be fully correlated between
different jet $R$ values but uncorrelated between different collision
centralities, so the uncertainties on the spectra were combined in
quadrature in evaluating \delRcp. The conservative assumption that the JER uncertainties are fully uncorrelated between
different centrality bins is based on the observation that the
differences between data and the HIJING MC sample in the fluctuation
analysis are not the same for all centralities.

\begin{figure*}[p]
\centerline{
\includegraphics[height=3.3in]{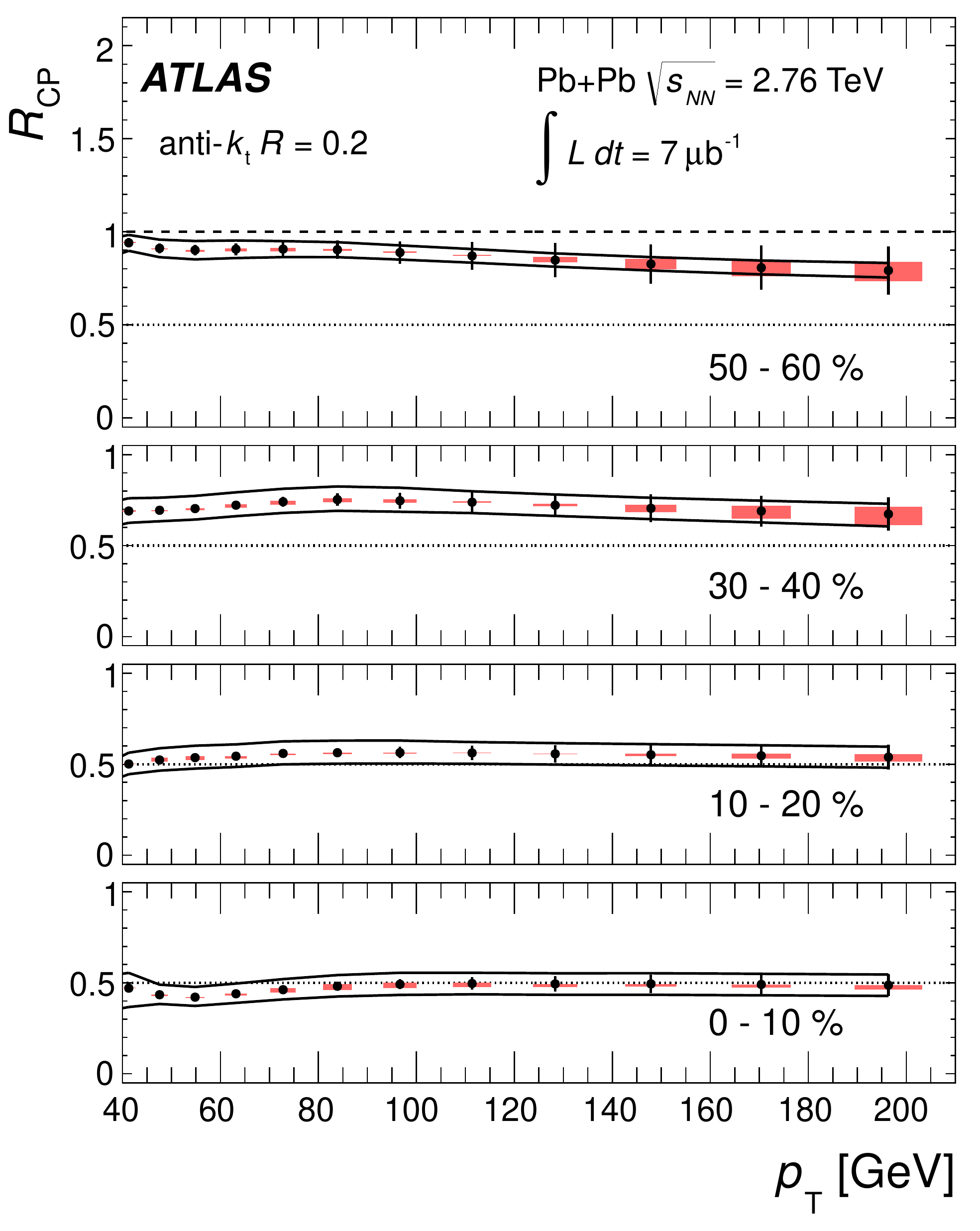}
\includegraphics[height=3.3in]{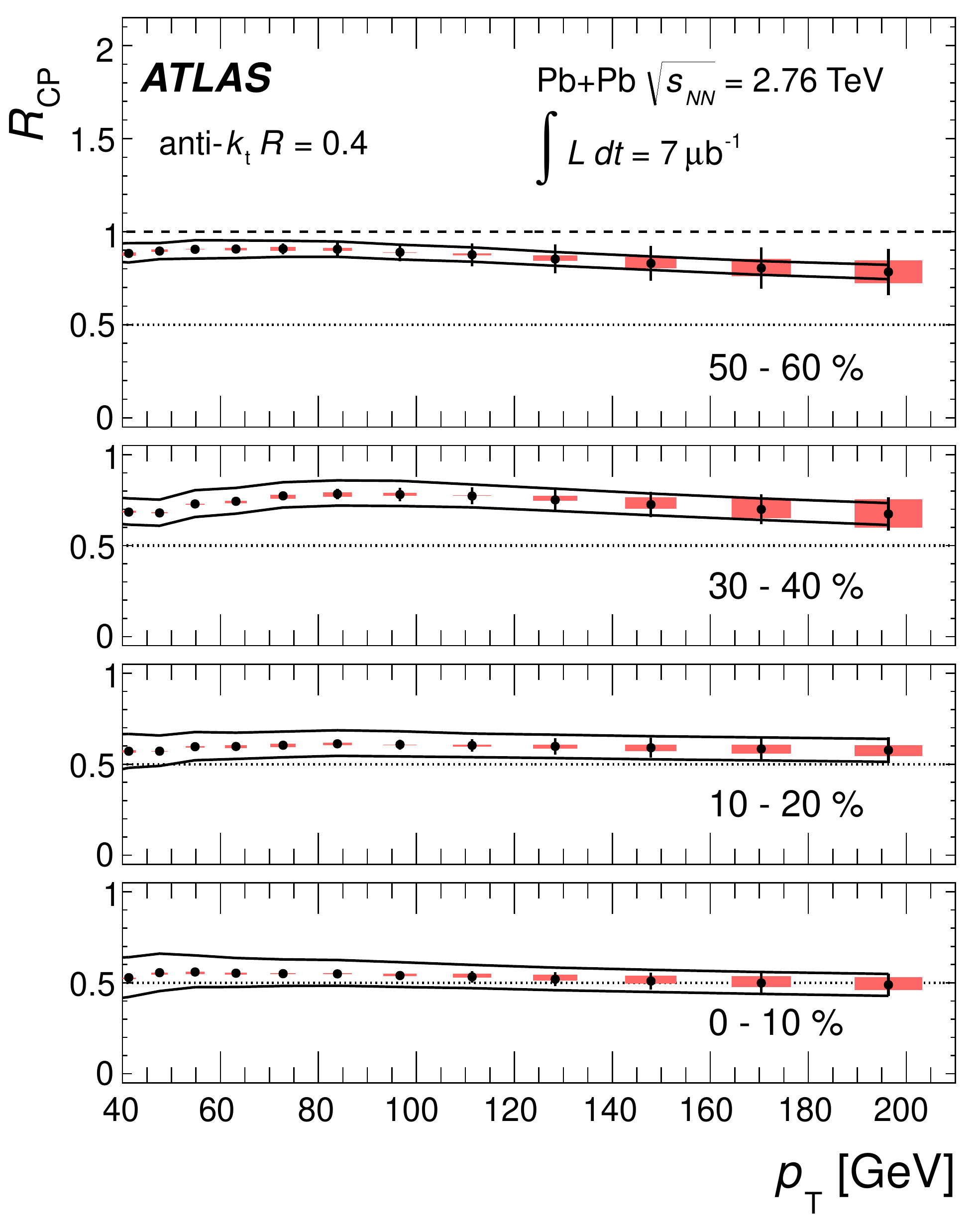}
}
\vspace{-0.15in}
\caption{
 \Rcp\ values as a function of jet
\pT\ for \RTwo\ (left) and \RFour\ (right) \antikt\ jets  
in four bins of collision centrality. The error bars indicate
statistical errors from the unfolding, the shaded boxes indicate
unfolding regularization systematic errors that are partially
correlated between points. The solid lines indicate  
systematic errors that are fully correlated between all points. 
The horizontal width of the systematic error band is chosen for
presentation purposes only. Dotted lines indicate $\Rcp =
0.5$, and the dashed lines on the top panels indicate $\Rcp = 1$. 
}
\label{fig:rcprfour}
\end{figure*}
The systematic uncertainties associated with the non-UE
contributions to the JER (described by the $a$ and $c$ terms in
Eq.~\ref{eq:resolparam}) were evaluated following procedures used by
ATLAS in previous \pp\ jet measurements  
\cite{Aad:2010ad}. New response matrices were generated by
applying an additional stochastic smearing to the \Dpt\ values,
and the systematic uncertainty was obtained by applying the procedure
described above. 

Systematic uncertainties on \Rcp\ due to the unfolding were evaluated by
changing the power index ($n$) in the functional form for \xini\ by $\pm
0.5$ and by varying the regularization parameter. The $\pm 0.5$ 
change in the power law index was chosen because it produces a
spectrum that changes relative to the default \xini\ over the measured
\pT\ range by a factor of about two -- the typical suppression
observed in central collisions. Thus, it covers the possibility that
the true \Rcp\ could increase to one or decrease to 0.25
over the measured \pT\ range.  To evaluate the potential
systematic uncertainty due to regularization, the unfolding was performed
with regularization parameters obtained from the fourth and sixth
singular values of the unfolding matrix, $\tau = s_4^2$ and $\tau =
s_6^2$. Systematic uncertainties on the spectra were determined from
the differences in the unfolded spectra. The resulting \delRcp\ values
were obtained assuming that the regularization 
uncertainties on the two spectra are uncorrelated.

The systematic uncertainty on the efficiency
correction was evaluated by comparing MC overlay and data overlay samples where
differences less than $5\%$ were observed on the ``turn on'' part of the
efficiency curve. A 5\% uncertainty due to the efficiency correction was applied
to \Rcp\ for $\pt < 100$~\GeV\ in the four most central bins.  To check for biases introduced by
the UE jet rejection, the analysis was repeated with a significantly weakened
rejection criterion in which jets were required to match a single track with
$\pt > 4$~\GeV. No significant differences in the \Rcp\ were observed
except for $\pt < 50$~\GeV\ where differences as high as 4\% were
found. These differences can be attributed to the contribution of additional UE jets.

The different contributions
to the total \delRcp\ are shown in Fig.~\ref{fig:syserror} for \RFour\ jets in the 0--10\%
centrality bin. The JES and \xini\ uncertainties are approximately
independent of \pt, while the JER  uncertainty
decreases with increasing \pt. The regularization uncertainty grows
with increasing \pt\ due to the poorer statistical precision of
the high-\pt\ points. The systematic uncertainties for the other radii
show similar \pt\ and centrality dependence, with the JES and JER
uncertainties increasing with jet radius as expected.

\begin{figure*}[p]
\centerline{
\includegraphics[height=2.75in]{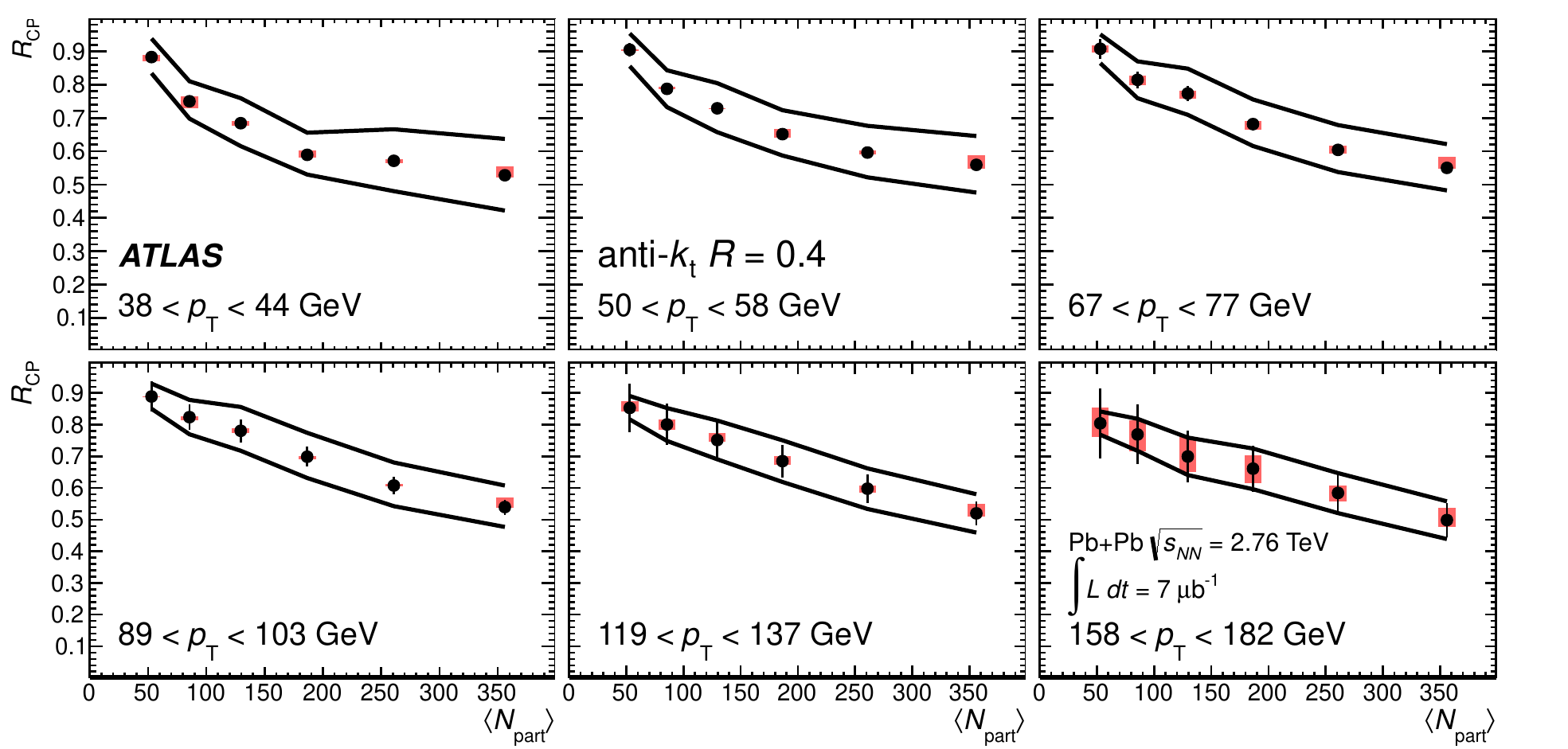}
}
\caption{
\Rcp\ values as a function of \Npart\
for \RFour\ \antikt\ jets in six \pT\ bins. \errordescr\
The horizontal errors indicate systematic uncertainties on \Npart. }
\label{fig:rcpcentdepend}
\end{figure*}
\begin{figure*}[]
\centerline{
\includegraphics[height=3.3in]{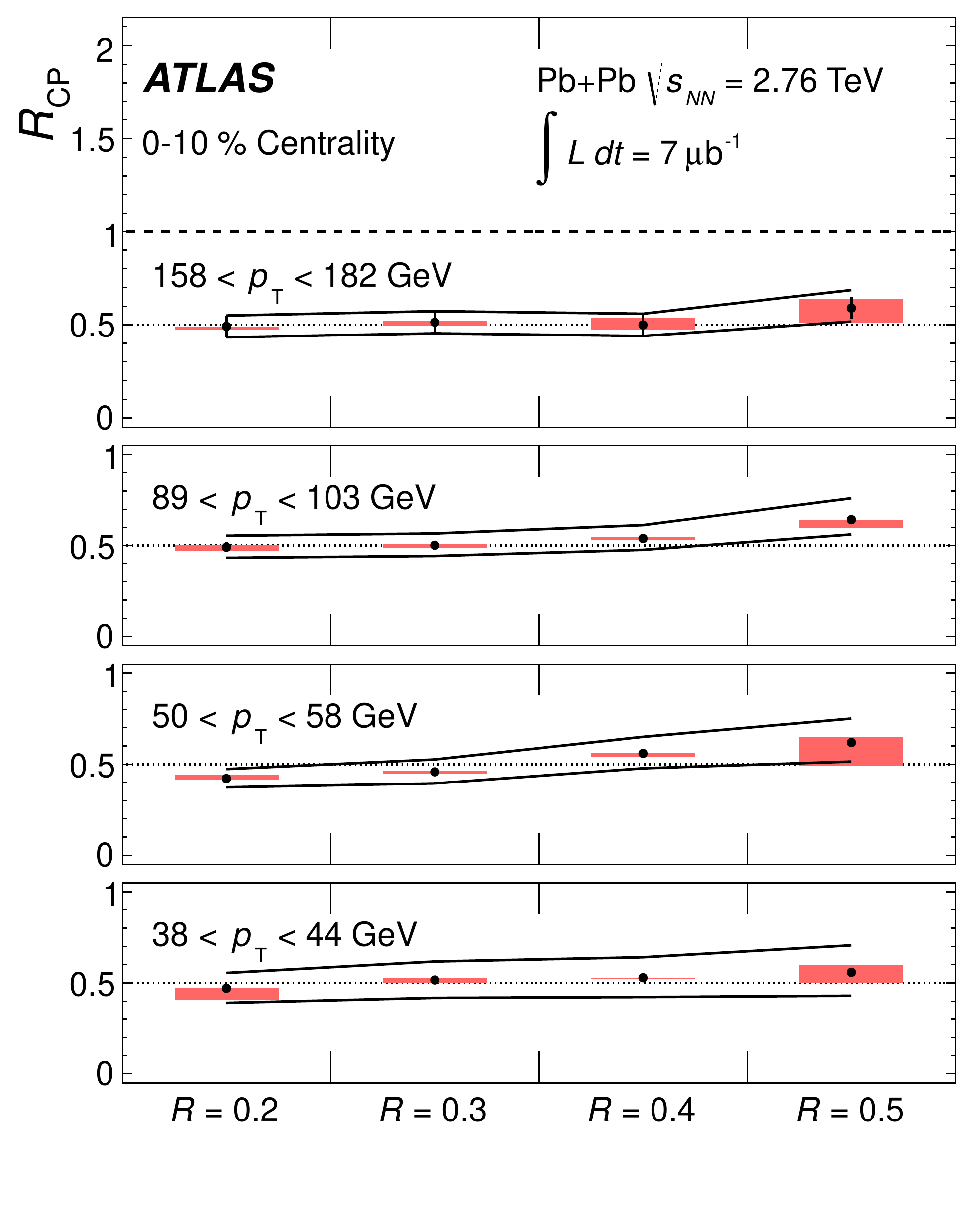}
\includegraphics[height=3.3in]{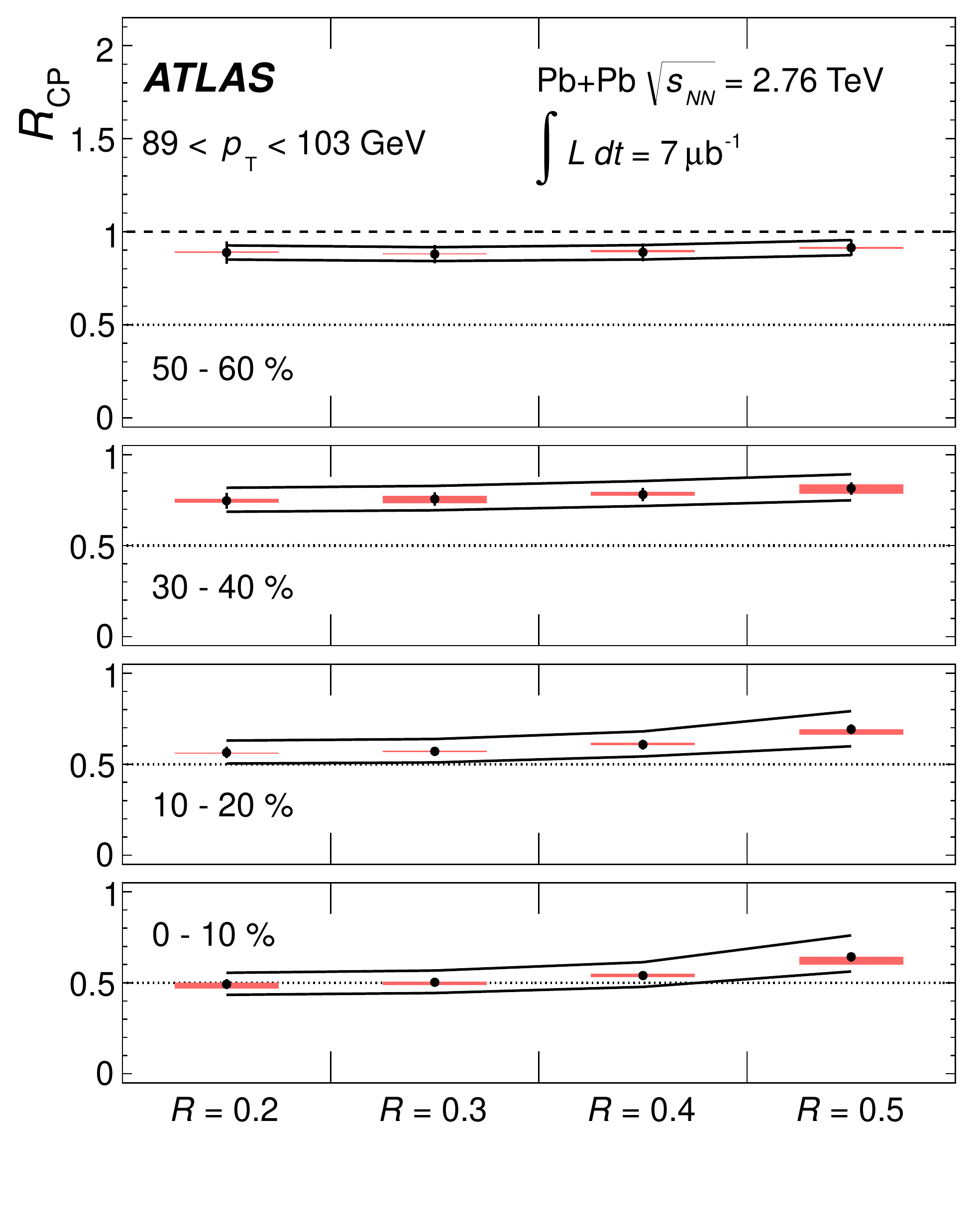}
}
\vspace{-0.25in}
\caption{
Left: \Rcp\ in the 0--10\% centrality bin as a function of jet radius
for four bins of jet \pT. Right: \Rcp\ as a function of jet
radius for four centrality bins for the \pT\ interval $89 < \pT < 103$~\GeV.
 \errordescr\  The horizontal width of the systematic error
band is chosen for presentation purposes only. Dotted lines indicate $\Rcp =
0.5$, and the dashed lines on the top panels indicate $\Rcp = 1$.}
\label{fig:rcprdepend}
\end{figure*}

\section{Results}
Figure~\ref{fig:rcprfour} shows the \Rcp\ values obtained for
\RTwo\ and \RFour\ jets as a function of \pT\ in four bins 
of collision centrality with three different error contributions:
statistical uncertainties, partially correlated systematic
uncertainties, and fully correlated uncertainties. 
The \Rcp\ values for all centralities and for both jet radii are
observed to have at most a weak variation with \pt. For the 0--10\%
centrality bin the \Rcp\ values for both jet radii show a factor
of about two suppression 
in the $1/\Ncoll$-scaled jet yield. For more peripheral collisions,
\Rcp\ increases at all jet \pT\ relative to central collisions,
with the \Rcp\ values reaching 0.9 
for the 50--60\% centrality bin. 
A more detailed evaluation of the centrality dependence of \Rcp\ for
\RFour\ jets is presented in Fig.~\ref{fig:rcpcentdepend}, which shows
\Rcp\ vs \Npart\ for six jet \pT\ bins. \Rcp\ decreases monotonically with increasing \Npart\ for all
\pT\ bins. The lower \pt\ bins, for which the data are more
statistically precise, show a variation of \Rcp\ with
\Npart\ that is most rapid at low \Npart. Trends similar to those shown in
Figs.~\ref{fig:rcprfour} and~\ref{fig:rcpcentdepend} are observed for all 
jet radii.
\begin{figure}
\centerline{
\includegraphics[width=0.49\textwidth]{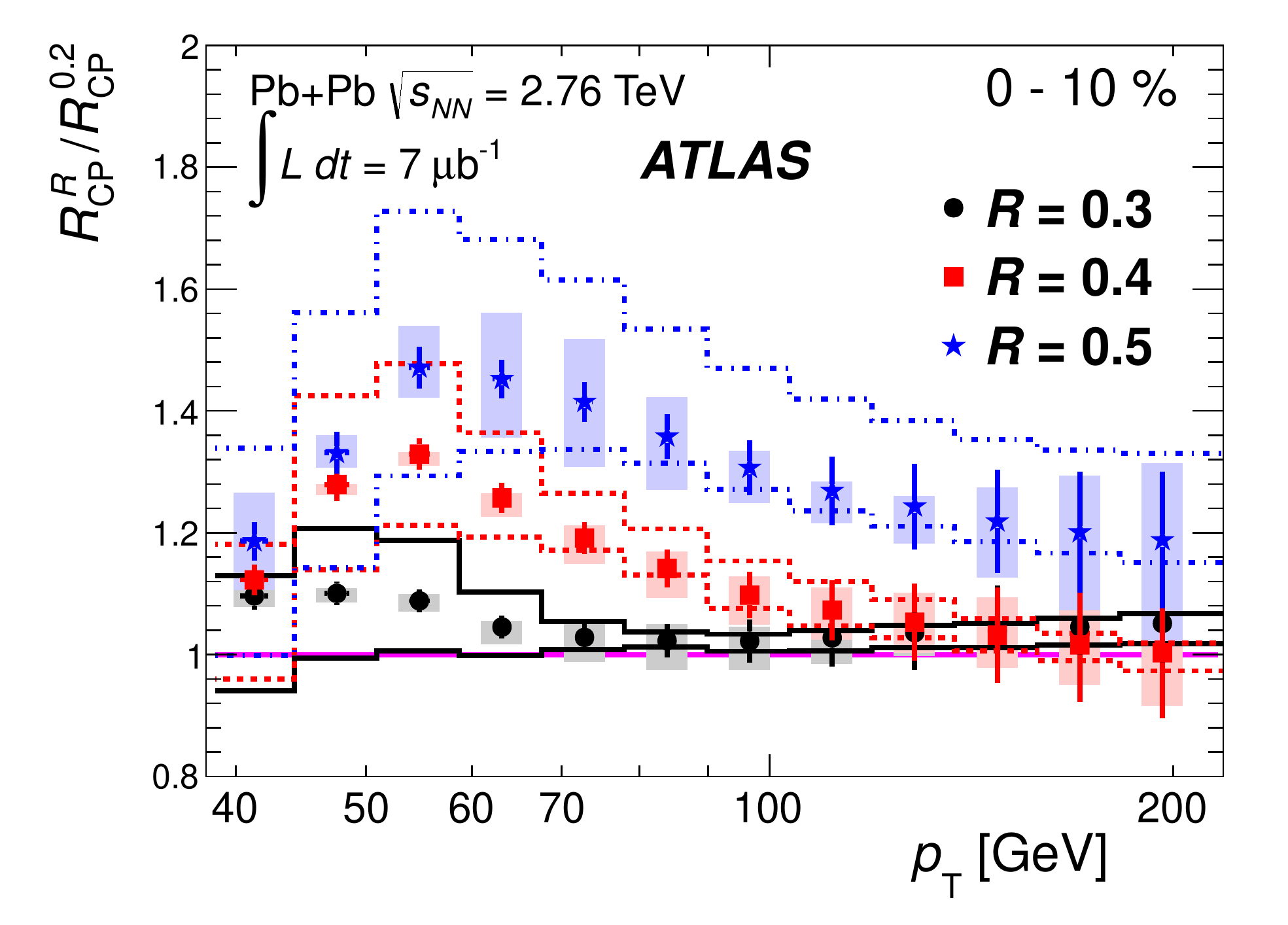}
}
\caption{
Ratios of \Rcp\ values between $R = 0.3, 0.4$ and 0.5 jets and $R =
0.2$ jets as a function of \pT\ in the 0--10\% centrality bin. The
error bars show statistical uncertainties (see text). The shaded boxes
indicate partially correlated systematic errors. The lines indicate
systematic errors that are fully correlated between different \pT\ bins.
}
\label{fig:rcpRratios}
\end{figure}

The dependence of \Rcp\ on jet radius is shown in
Fig.~\ref{fig:rcprdepend} for the 0--10\% centrality bin
in four jet \pT\ intervals (left) and for different centrality
bins in the $89 < \pT < 103$~\GeV\ \ bin (right). For
this figure, the shaded boxes indicate the combined contribution of
systematic uncertainties due to regularization, \xini, and
efficiency, which are only partially correlated between points. All
other systematic errors are fully correlated and are indicated by
solid lines. The results in Fig.~\ref{fig:rcprdepend} show a weak
variation of \Rcp\ with $R$, that is nonetheless significant when
taking into account the correlations in the errors between the
different $R$ values.

To demonstrate this conclusion more clearly,
Fig.~\ref{fig:rcpRratios} shows the ratio of \Rcp\ values between $R =
0.3, 0.4$ and 0.5 jets and $R = 0.2$ jets, \RcpRatio, as a function of
\pT\ for the 0--10\% centrality bin. When evaluating the ratio, there
is significant cancellation between the correlated
systematic uncertainties. Statistical correlations between the jet yields for the different radii
were evaluated in the measured spectra and  tracked through the
unfolding procedure separately for the 0-10\% and 60-80\% centrality
bins. Those correlations were then included when evaluating the
statistical errors on \RcpRatio\ shown in Fig.~\ref{fig:rcpRratios}.
The results in that figure indicate a significant dependence of
\Rcp\ on jet radius. For $\pT < 100$~\GeV\, the
\RcpRatio\ values for both \RFour\ and \RFive\ differ from one beyond
the statistical and systematic uncertainties. The deviation persists
for \RFive\ above 100~\GeV. A similar, but weaker dependence is  
observed in the 10--20\% centrality bin. In more peripheral bins, no
significant radial dependence is observed. The differences between
\Rcp\ values for the different jet radii
increase with decreasing \pT, except for the lowest two \pT\ bins.
However, direct comparisons of \Rcp\ between different jet radii at
low \pT\ should be treated with care as the same jets measured
using smaller radii will tend to appear in lower \pT\ bins than when
measured with a larger radius.  

\section{Conclusions}
This Letter presents results of measurements of the centrality
dependence of jet suppression, characterized by the inclusive jet central-to-peripheral
ratio, \Rcp, in \PbPb\ collisions at 2.76~\TeV\ per nucleon at the LHC.
The measurements were performed over the \pt\ range $38 < \pt < 210$~\GeV\ for
anti-\kt\ jets of radii $R = 0.2, 0.3, 0.4$ and $0.5$. The inclusive jet yield is
observed to be suppressed by a factor of about two in central
collisions relative to peripheral collisions with at most a weak
\pt\ dependence to the suppression. The suppression varies
monotonically with collision centrality over the measured \pT\ range
and for all jet radii. The \Rcp\ at fixed \pT\ is observed to vary with
jet radius increasing gradually from \RTwo\ to \RFive. That variation
is most significant for $\pT < 100$~\GeV\ where more than a 30\%
variation is observed. These results provide the
first direct measurement of inclusive jet suppression in heavy ion
collisions. The substantial suppression of the jet yield observed at
all \pT\ values complements the previous measurements of dijet
transverse energy imbalance in \PbPb\ collisions 
at the LHC~\cite{Aad:2010bu, Chatrchyan:1327643,CMS:2012ni}.

\section*{Acknowledgments}
We thank CERN for the very successful operation of the LHC, as well as the
support staff from our institutions without whom ATLAS could not be
operated efficiently.

We acknowledge the support of ANPCyT, Argentina; YerPhI, Armenia; ARC,
Australia; BMWF, Austria; ANAS, Azerbaijan; SSTC, Belarus; CNPq and FAPESP,
Brazil; NSERC, NRC and CFI, Canada; CERN; CONICYT, Chile; CAS, MOST and NSFC,
China; COLCIENCIAS, Colombia; MSMT CR, MPO CR and VSC CR, Czech Republic;
DNRF, DNSRC and Lundbeck Foundation, Denmark; EPLANET and ERC, European Union;
IN2P3-CNRS, CEA-DSM/IRFU, France; GNAS, Georgia; BMBF, DFG, HGF, MPG and AvH
Foundation, Germany; GSRT, Greece; ISF, MINERVA, GIF, DIP and Benoziyo Center,
Israel; INFN, Italy; MEXT and JSPS, Japan; CNRST, Morocco; FOM and NWO,
Netherlands; RCN, Norway; MNiSW, Poland; GRICES and FCT, Portugal; MERYS
(MECTS), Romania; MES of Russia and ROSATOM, Russian Federation; JINR; MSTD,
Serbia; MSSR, Slovakia; ARRS and MVZT, Slovenia; DST/NRF, South Africa;
MICINN, Spain; SRC and Wallenberg Foundation, Sweden; SER, SNSF and Cantons of
Bern and Geneva, Switzerland; NSC, Taiwan; TAEK, Turkey; STFC, the Royal
Society and Leverhulme Trust, United Kingdom; DOE and NSF, United States of
America.

The crucial computing support from all WLCG partners is acknowledged
gratefully, in particular from CERN and the ATLAS Tier-1 facilities at
TRIUMF (Canada), NDGF (Denmark, Norway, Sweden), CC-IN2P3 (France),
KIT/GridKA (Germany), INFN-CNAF (Italy), NL-T1 (Netherlands), PIC (Spain),
ASGC (Taiwan), RAL (UK) and BNL (USA) and in the Tier-2 facilities
worldwide.

\bibliography{JetSuppressionRcpLetter}

\begin{thebibliography}{10}
\expandafter\ifx\csname url\endcsname\relax
  \def\url#1{\texttt{#1}}\fi
\expandafter\ifx\csname urlprefix\endcsname\relax\def\urlprefix{URL }\fi
\expandafter\ifx\csname href\endcsname\relax
  \def\href#1#2{#2} \def\path#1{#1}\fi

\bibitem{Abreu:2007kv}
{Armesto, N. (Ed.)}.
\newblock J.~Phys.~G 35 (2008) 054001.
\newblock \href {http://arxiv.org/abs/0711.0974} {\path{arXiv:0711.0974}}.

\bibitem{Wang:1994fx}
X.-N. Wang, M.~Gyulassy, M.~Plumer.
\newblock Phys.~Rev. D~51 (1995) 3436--3446.
\newblock \href {http://arxiv.org/abs/hep-ph/9408344}
  {\path{arXiv:hep-ph/9408344}}.

\bibitem{Baier:1998yf}
R.~Baier, Y.~L. Dokshitzer, A.~H. Mueller, D.~Schiff.
\newblock Phys.~Rev. C~58 (1998) 1706--1713.
\newblock \href {http://arxiv.org/abs/hep-ph/9803473}
  {\path{arXiv:hep-ph/9803473}}.

\bibitem{Gyulassy:2000fs}
M.~Gyulassy, P.~Levai, I.~Vitev.
\newblock Phys.~Rev.~Lett. 85 (2000) 5535--5538.
\newblock \href {http://arxiv.org/abs/nucl-th/0005032}
  {\path{arXiv:nucl-th/0005032}}.

\bibitem{Adler:2003qi}
S.~S. Adler, et~al.
\newblock Phys.~Rev.~Lett. 91 (2003) 072301.
\newblock \href {http://arxiv.org/abs/nucl-ex/0304022}
  {\path{arXiv:nucl-ex/0304022}}.

\bibitem{Adams:2003kv}
J.~Adams, et~al.
\newblock Phys.~Rev.~Lett. 91 (2003) 172302.
\newblock \href {http://arxiv.org/abs/nucl-ex/0305015}
  {\path{arXiv:nucl-ex/0305015}}.

\bibitem{Arsene:2003yk}
I.~Arsene, et~al.
\newblock Phys.~Rev.~Lett. 91 (2003) 072305.
\newblock \href {http://arxiv.org/abs/nucl-ex/0307003}
  {\path{arXiv:nucl-ex/0307003}}.

\bibitem{Back:2004ra}
B.~B. Back, et~al.
\newblock Phys.~Rev.~Lett. 94 (2005) 082304.
\newblock \href {http://arxiv.org/abs/nucl-ex/0405003}
  {\path{arXiv:nucl-ex/0405003}}.

\bibitem{Adler:2002tq}
C.~Adler, et~al.
\newblock Phys.~Rev.~Lett. 90 (2003) 082302.
\newblock \href {http://arxiv.org/abs/nucl-ex/0210033}
  {\path{arXiv:nucl-ex/0210033}}.

\bibitem{phenix:2008cqb}
A.~Adare, et~al.
\newblock Phys.~Rev. C~78 (2008) 014901.
\newblock \href {http://arxiv.org/abs/0801.4545} {\path{arXiv:0801.4545}}.

\bibitem{Adcox:2004mh}
K.~Adcox, et~al.
\newblock Nucl.~Phys. A~757 (2005) 184--283.
\newblock \href {http://arxiv.org/abs/nucl-ex/0410003}
  {\path{arXiv:nucl-ex/0410003}}.

\bibitem{Adams:2005dq}
J.~Adams, et~al.
\newblock Nucl.~Phys. A~757 (2005) 102--183.
\newblock \href {http://arxiv.org/abs/nucl-ex/0501009}
  {\path{arXiv:nucl-ex/0501009}}.

\bibitem{Aad:2010bu}
{ATLAS Collaboration}.
\newblock Phys.~Rev.~Lett. 105 (2010) 252303.
\newblock \href {http://arxiv.org/abs/1011.6182} {\path{arXiv:1011.6182}}.

\bibitem{Chatrchyan:1327643}
{CMS Collaboration}.
\newblock Phys.~Rev.~C 84 (2011) 024906.
\newblock \href {http://arxiv.org/abs/1102.1957} {\path{arXiv:1102.1957}}.

\bibitem{CMS:2012ni}
{CMS Collaboration}, {CERN-PH-EP-2012-042} (2012).
\newblock \href {http://arxiv.org/abs/1202.5022} {\path{arXiv:1202.5022}}.

\bibitem{Bjorken:1982tu}
J.~D. Bjorken, Fermilab Preprint, {FERMILAB-PUB-82-059-THY}.

\bibitem{Armesto:2011ht}
N.~Armesto, B.~Cole, C.~Gale, W.~A. Horowitz, P.~Jacobs, et~al., submitted to
  Phys.~Rev.~C.
\newblock \href {http://arxiv.org/abs/1106.1106} {\path{arXiv:1106.1106}}.

\bibitem{Vitev:2008rz}
I.~Vitev, S.~Wicks, B.-W. Zhang.
\newblock JHEP 0811 (2008) 093.
\newblock \href {http://arxiv.org/abs/0810.2807} {\path{arXiv:0810.2807}}.

\bibitem{Vitev:2009rd}
I.~Vitev, B.-W. Zhang.
\newblock Phys.~Rev.~Lett. 104 (2010) 132001.
\newblock \href {http://arxiv.org/abs/0910.1090} {\path{arXiv:0910.1090}}.

\bibitem{He:2011pd}
Y.~He, I.~Vitev, B.-W. Zhang.
\newblock \href {http://arxiv.org/abs/1105.2566} {\path{arXiv:1105.2566}}.

\bibitem{Qin:2010mn}
G.-Y. Qin, B.~Muller.
\newblock Phys.Rev.Lett. 106 (2011) 162302.
\newblock \href {http://arxiv.org/abs/1012.5280} {\path{arXiv:1012.5280}},
  \href {http://dx.doi.org/10.1103/PhysRevLett.108.189904,
  10.1103/PhysRevLett.106.162302} {\path{doi:10.1103/PhysRevLett.108.189904,
  10.1103/PhysRevLett.106.162302}}.

\bibitem{Cacciari:2008gp}
M.~Cacciari, G.~P. Salam, G.~Soyez.
\newblock JHEP 0804 (2008) 063.
\newblock \href {http://arxiv.org/abs/0802.1189} {\path{arXiv:0802.1189}}.

\bibitem{Aad:2008zzm}
{ATLAS Collaboration}.
\newblock JINST 3 (2008) S08003.

\bibitem{Aad:2010bx}
{ATLAS Collaboration}.
\newblock Eur.~Phys.~J. C~70 (2010) 787--821.
\newblock \href {http://arxiv.org/abs/1004.5293} {\path{arXiv:1004.5293}}.

\bibitem{Cornelissen:2008zzc}
T.~Cornelissen, M.~Elsing, I.~Gavrilenko, W.~Liebig, E.~Moyse, et~al.
\newblock J.~Phys.~Conf.~Ser. 119 (2008) 032014.

\bibitem{ATLAS:2011ah}
{ATLAS Collaboration}.
\newblock Phys.~Lett.~ B~707 (2012) 330--348.
\newblock \href {http://arxiv.org/abs/1108.6018} {\path{arXiv:1108.6018}}.

\bibitem{Djuvsland:2010qs}
O.~Djuvsland, J.~Nystrand.
\newblock Phys.~Rev. C~83 (2010) 041901.
\newblock \href {http://arxiv.org/abs/1011.4908} {\path{arXiv:1011.4908}}.

\bibitem{Alver:2008aq}
B.~Alver, M.~Baker, C.~Loizides, P.~Steinberg.
\newblock \href {http://arxiv.org/abs/0805.4411} {\path{arXiv:0805.4411}}.

\bibitem{Miller:2007ri}
M.~L. Miller, K.~Reygers, S.~J. Sanders, P.~Steinberg.
\newblock Ann.~Rev.~Nucl.~Part.~Sci. 57 (2007) 205--243.
\newblock \href {http://arxiv.org/abs/nucl-ex/0701025}
  {\path{arXiv:nucl-ex/0701025}}.

\bibitem{atlassim}
{ATLAS Collaboration}.
\newblock Eur.~Phys.~J. C~70 (2010) 823--874.
\newblock \href {http://arxiv.org/abs/1005.4568} {\path{arXiv:1005.4568}}.

\bibitem{Wang:1991hta}
X.-N. Wang, M.~Gyulassy.
\newblock Phys.~Rev. D~44 (1991) 3501--3516.

\bibitem{Masera:2009zz}
M.~Masera, G.~Ortona, M.~Poghosyan, F.~Prino.
\newblock Phys.~Rev. C~79 (2009) 064909.

\bibitem{Agostinelli:2002hh}
S.~Agostinelli, et~al.
\newblock Nucl.~Instrum.~Meth. A~506 (2003) 250--303.

\bibitem{:2010ir}
{ATLAS Collaboration}.
\newblock New~J.~Phys. 13 (2010) 053033.
\newblock \href {http://arxiv.org/abs/1012.5104} {\path{arXiv:1012.5104}}.

\bibitem{Sjostrand:2006za}
T.~Sjostrand, S.~Mrenna, P.~Z. Skands.
\newblock JHEP 0605 (2006) 026.
\newblock \href {http://arxiv.org/abs/hep-ph/0603175}
  {\path{arXiv:hep-ph/0603175}}.

\bibitem{Aad:2011he}
{ATLAS Collaboration}, submitted to Eur.~Phys.~J.~C.
\newblock \href {http://arxiv.org/abs/1112.6426} {\path{arXiv:1112.6426}}.

\bibitem{Poskanzer:1998yz}
A.~M. Poskanzer, S.~Voloshin.
\newblock Phys.~Rev. C~58 (1998) 1671--1678.
\newblock \href {http://arxiv.org/abs/nucl-ex/9805001}
  {\path{arXiv:nucl-ex/9805001}}.

\bibitem{Aad:2009wy}
{ATLAS Collaboration}, {CERN-OPEN-2008-20}.
\newblock \href {http://arxiv.org/abs/0901.0512} {\path{arXiv:0901.0512}}.

\bibitem{Hocker:1995kb}
A.~Hocker, V.~Kartvelishvili.
\newblock Nucl.~Instrum.~Meth. A~372 (1996) 469--481.
\newblock \href {http://arxiv.org/abs/hep-ph/9509307}
  {\path{arXiv:hep-ph/9509307}}.

\bibitem{Adye:2011gm}
T.~Adye, {Proceedings of the PHYSTAT 2009 Workshop, CERN, Geneva, Switzerland,
  January 2011, CERN-2011-006, pp 313}.
\newblock \href {http://arxiv.org/abs/1105.1160} {\path{arXiv:1105.1160}}.

\bibitem{Aad:2010ad}
{Atlas Collaboration}.
\newblock Eur.~Phys.~J. C71 (2011) 1512.
\newblock \href {http://arxiv.org/abs/1009.5908} {\path{arXiv:1009.5908}}.

\end{thebibliography}
\bibliographystyle{elsarticle-num}

\clearpage
\onecolumn
\begin{flushleft}
{\Large The ATLAS Collaboration}

\bigskip

G.~Aad$^{\rm 48}$,
B.~Abbott$^{\rm 111}$,
J.~Abdallah$^{\rm 11}$,
S.~Abdel~Khalek$^{\rm 115}$,
A.A.~Abdelalim$^{\rm 49}$,
O.~Abdinov$^{\rm 10}$,
B.~Abi$^{\rm 112}$,
O.S.~AbouZeid$^{\rm 158}$,
H.~Abramowicz$^{\rm 153}$,
H.~Abreu$^{\rm 136}$
% ...
\bigskip

% ...
$^{10}$ Institute of Physics, Azerbaijan Academy of Sciences, Baku, Azerbaijan\\
$^{11}$ Institut de F\'isica d'Altes Energies and Departament de F\'isica de la Universitat Aut\`onoma  de Barcelona and ICREA, Barcelona, Spain\\
% ...
$^{48}$ Fakult\"{a}t f\"{u}r Mathematik und Physik, Albert-Ludwigs-Universit\"{a}t, Freiburg i.Br., Germany\\
$^{49}$ Section de Physique, Universit\'e de Gen\`eve, Geneva, Switzerland\\
% ...
$^{111}$ Homer L. Dodge Department of Physics and Astronomy, University of Oklahoma, Norman OK, United States of America\\
$^{112}$ Department of Physics, Oklahoma State University, Stillwater OK, United States of America\\
% ...
$^{115}$ LAL, Univ. Paris-Sud and CNRS/IN2P3, Orsay, France\\
% ...
$^{136}$ DSM/IRFU (Institut de Recherches sur les Lois Fondamentales de l'Univers), CEA Saclay (Commissariat a l'Energie Atomique), Gif-sur-Yvette, France\\
% ...
$^{153}$ Raymond and Beverly Sackler School of Physics and Astronomy, Tel Aviv University, Tel Aviv, Israel\\
% ...
$^{158}$ Department of Physics, University of Toronto, Toronto ON, Canada\\
% ...
\end{flushleft}

\end{document}